\documentclass[preprint,12pt,authoryear]{elsarticle}
\usepackage[margin=2.5cm]{geometry}
\usepackage[hidelinks]{hyperref}
\usepackage{siunitx}
\usepackage{bm}
\usepackage{enumitem}
\usepackage{xcolor}

\usepackage{soul}
\usepackage{nomencl}
\usepackage{multicol}
\usepackage{framed}
\usepackage[most]{tcolorbox}
\makenomenclature
\usepackage[normalem]{ulem}

\usepackage{amssymb}
\usepackage{amsmath}
\usepackage{esint}

\begin{document}

\begin{frontmatter}

\title{A predictive model for bubble--particle collisions in turbulence}

\author[UT]{Timothy T.K. Chan\fnref{fn1}} 
\author[UT,Aachen]{Linfeng Jiang\fnref{fn1}\corref{cor1}} 
\ead{l.jiang@aia.rwth-aachen.de}
\author[UT,Aachen]{Dominik Krug} 

\cortext[cor1]{Corresponding author}
\fntext[fn1]{T.T.K. Chan and L. Jiang contributed equally to this work.}

\address[UT]{Physics of Fluids Group, Max Planck Center for Complex Fluid Dynamics, and J. M. Burgers Centre for Fluid Mechanics, University of Twente, P.O. Box 217, 7500 AE Enschede, The Netherlands}
\address[Aachen]{Institute of Aerodynamics, RWTH Aachen University, Wüllnerstraße 5a, 52062 Aachen, Germany}

\begin{abstract}
The modelling of bubble--particle collisions is crucial to improving the efficiency of industrial processes such as froth flotation. Although such systems usually have turbulent flows and the bubbles are typically much larger than the particles, there currently exist no predictive models for this case which consistently include finite-size effects in the interaction with the bubbles as well as inertial effects for the particles simultaneously. As a first step, \citet{jiang_how_2024} proposed a frozen turbulence approach which captures the collision rate between finite-size bubbles and inertial particles in homogeneous isotropic turbulence using the bubble slip velocity probability density function measured from simulations as an input. In this study, we further develop this approach into a model where the bubble--particle collision rate can be predicted \textit{a priori} based on the bubble, particle, and turbulence properties. By comparing the predicted collision rate with simulations of bubbles with Stokes numbers of 2.8 and 6.3, and particles with Stokes numbers ranging from 0.01 to 2 in turbulence with a Taylor Reynolds number of 64, good agreement is found between model and simulations for Froude number $Fr \leq 0.25$. Beyond this range of bubble Stokes number, we propose a criterion for using our model and discuss the model's validity. Evaluating our model at typical flotation parameters indicates that particle inertia {and settling} effects are usually important. Generally, smaller bubbles, larger particles, and stronger turbulence increase the overall collision rate.

\end{abstract}

\begin{graphicalabstract}
\includegraphics[width = \linewidth]{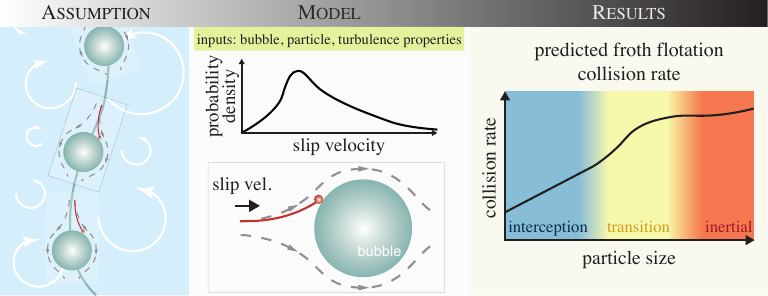}
\end{graphicalabstract}

\begin{highlights}
\item {Developed a fully predictive model for the bubble-particle collision rate in homogeneous isotropic turbulence}
\item {Incorporates particle inertia and bubble-induced flow field distortion into the collision model}
\item {Requires only bubble, particle, and turbulence properties as model inputs}
\item {Provides a potential subgrid-scale model to support industrial-scale froth flotation simulations}
\item {Validated through direct numerical simulations of finite-size bubbles and point-particles in turbulent flows}
\end{highlights}


\end{frontmatter}

\section{Introduction}\label{sec:intro_FS}
Bubble--particle collisions lie at the heart of froth flotation --- an industrial process which selectively recovers particles based on their hydrophobicity with applications ranging from mineral extraction to wastewater treatment \citep{rubio_overview_2002,nguyen_colloidal_2004}. In this process, bubbles and particles are injected into a cell filled with water so they can collide with each other. As the bubbles and particles collide, the target particles (which are hydrophobic) attach to the bubbles and are floated to the surface, where they can be removed. In other words, every single particle has to collide with a bubble before it is recovered. Given the broad range of application of froth flotation, as well as the large size of flotation cells and the complex flows therein which precludes particle-resolved simulations, there is a strong incentive to model the bubble--particle collision rate.
The collision rate per volume $Z$ can be expressed as:
\begin{equation}    \label{eq:collKRate}
    Z = \Gamma n_bn_p,
\end{equation}
where $n_p$ and $n_b$ are the particle and bubble number density. The collision kernel $\Gamma$ serves as a fundamental parameter in quantifying the frequency at which bubbles and particles collide per unit volume and time.

In quiescent fluid, the bubble–particle collision process is primarily deterministic, governed by hydrodynamic mechanisms such as interception, gravitational sedimentation, and inertia. Particles approach rising bubbles along trajectories shaped by these mechanisms, and the collision kernel in such systems can be equivalently quantified by the collision efficiency ($E_c$). The collision efficiency quantifies the probability of collision, defined as the ratio of the number of particles that actually collide with the bubble to the total number of particles encountered along the bubble’s trajectory.
The dominant collision mechanism depends strongly on particle properties, notably density $\rho_p$ and radius $r_p$, relative to those of the bubble. Typically, particles are significantly smaller than bubbles, making the local flow field around the bubble crucial for the collision dynamics. In the low-inertia limit, particles closely follow fluid streamlines, resulting in interception-driven collisions. The interceptional collision efficiency can be analytically derived based on the flow solutions around a bubble, as presented by \citet{weber_interceptional_1983,YOON02021989, sutherland1948physical}, and describes how streamlines bring particles into contact with the bubble surface.
When particle inertia increases, particles deviate from streamlines and collide due to inertial effects as well, leading to a marked rise in collision efficiency (\citet{schulze_hydrodynamics_1989}). Additionally, particle gravitational settling can enhance collision rates, especially for dense or large particles, with corresponding models extending the efficiency predictions (\citet{RALSTON1999207,weber_interceptional_1983}).
Intuitively, the contributions of interception, inertia, and gravity to the total collision efficiency may be considered additive (\citet{schulze_hydrodynamics_1989}). More recently, \citet{huang_effect_2012} studied the impact of the interface contamination on the collision efficiency by means of direct numerical simulations.

While the deterministic mechanisms governing bubble–particle collisions in quiescent fluid provide valuable foundational insights, real flotation systems typically operate under turbulent conditions where flow unsteadiness and velocity fluctuations profoundly influence collision dynamics. Unlike the relatively predictable trajectories in static environments, turbulence introduces stochastic relative motions and enhanced mixing that can significantly increase collision rates beyond those captured by quiescent fluid models. To this end, more parameters are relevant in determining the collision kernel in turbulence. The most relevant ones are the Taylor Reynolds number $Re_\lambda$, which indicates the turbulence intensity, the Froude number $Fr$ characterising the strength of turbulence relative to gravity, and the bubble and particle Stokes numbers based on the Kolmogorov time scale $St_b$ and $St_p$. These are defined as follows:
\begin{equation}
    Re_\lambda = \sqrt{\frac{15}{\nu\varepsilon}}u'^2,\, Fr = \frac{a_\eta}{\mathfrak{g}} = \frac{u_\eta}{\mathfrak{g}\tau_\eta},\, St_b = \frac{\tau_b}{\tau_\eta},\, St_p = \frac{\tau_p}{\tau_\eta},
\end{equation}
where $u'$ is the single-component root-mean-square (r.m.s.) fluid velocity, $\nu$ is the kinematic viscosity, $\varepsilon$ is the average rate of turbulent dissipation, $\mathfrak{g}$ is the gravitational acceleration, $\tau_b = r_b^2(2\rho_b/\rho_f + 1)/(9\nu)$ is the bubble response time, $\tau_p = r_p^2(2\rho_p/\rho_f + 1)/(9\nu)$ is the particle response time, $r_b$ is the bubble radius, $\rho_b$ is the bubble density, $\rho_f$ is the fluid density, and $\tau_\eta = \sqrt{\nu/\varepsilon}$ and $u_\eta = (\nu\varepsilon)^{0.25}$ are the Kolmogorov time scale and velocity scale, respectively. The particle response time characterizes the time scale over which a particle adjusts its velocity to match changes in the surrounding fluid velocity. It can be derived by nondimensionalizing the equation of motion for point particles subjected to fluid forces, as detailed in seminal works \citep{maxey_equation_1983, gatignol_faxen_1983}. In turbulent flows, this formulation is generally valid for particles that are small compared to the Kolmogorov length scale $\eta$.

Modeling the bubble–particle collision kernel fundamentally depends on a detailed understanding of both bubble and particle dynamics within turbulent flows. Early models for bubble–particle collisions have naturally drawn upon the extensive literature on particle–particle collisions (see for example \citet{pumir_collisional_2016}), where dynamics are often described by point-particle approximations. One of the seminal works in this area is the model proposed by \citet{saffman_collision_1956}, which accounts for shear-induced collisions between tracer-like particles in turbulence. In contrast, \citet{abrahamson_collision_1975} developed a collision kernel model applicable to particles with very large inertia, characterizing collision rates driven primarily by relative inertial motion.
Practically, flotation systems operate with bubbles and particles that lie in an intermediate inertia regime. Addressing this, \citet{yuu_collision_1984} proposed a collision model that incorporates both shear and inertia induced relative motions for particles with intermediate Stokes numbers, thus bridging the gap between the two limiting cases.
Nevertheless, these models predominantly quantify the “geometric collision” rate typical of particle–particle interactions (\citet{chan_bubbleparticle_2023} and \citet{chan_effect_2024} for an overview).
Insights from studies on bubble–particle collision efficiency in quiescent fluids clearly demonstrate that disregarding the pronounced size difference between bubbles and particles introduces significant inaccuracies in collision kernel estimates. The large size disparity indicates that the flow disturbance created by the bubble strongly influences particle trajectories, fundamentally altering collision probabilities compared to particle–particle interactions. 
However, accounting for the effect of turbulence in addition to the flow distortion by the bubble is a long standing problem in the literature \citep{pyke_bubble_2003,nguyen2016,wang_review_2018,hassanzadeh_review_2018,wang_development_2020,wang_effect_2022}. One approach has been to estimate the collision kernel in turbulent flows by multiplying the geometric collision kernel by the collision efficiency (\citet{pyke_bubble_2003,koh_cfd_2006}). 
However, this approach is conceptually problematic as the instantaneous and averaged values of $Re_b$ can differ significantly. Additionally, \citet{pyke_bubble_2003} incorrectly defines $Re_b$ with the bubble rise velocity instead of the bubble slip velocity.

In contrast, the recent model by \citet{kostoglou_generalized_2020} {takes a different approach and }implicitly includes the collision efficiency concept in turbulence in a consistent manner. This is done by directly evaluating the particle influx around an isolated bubble that is exposed to an unsteady incident flow whose fluctuations are determined by the bubble r.m.s. slip velocity, and by incorporating the distorted flow field around the bubble using the expression given by \citet{nguyen_colloidal_2004}. In addition, they consider the collisions caused by small scale turbulence. Their model is proposed for the case of uniformly distributed tracer-like particles colliding with bubbles, so it has limited applicability in real-life, where inertial effects are significant for most particles. 
More recently, \citet{jiang_how_2024} developed an approach applicable to inertial, one-way coupled particles by decomposing the bubble trajectory into small segments over which the bubble slip velocity is assumed constant. This reduces the instantaneous collision process to a deterministic quiescent fluid scenario with a slip velocity equal to the instantaneous value. However, in their study the bubble slip velocity distribution is not predicted, but it is obtained directly from numerical simulations. As a result, the model remains unclosed in terms of collision kernel prediction. The present work builds upon the framework of \citet{jiang_how_2024} by developing a complete model for the slip velocity probability distribution, thereby enabling standalone prediction of the bubble–particle collision kernel without reliance on additional simulation data.

This paper is organised as follows: {We first We first present our model in \S\ref{sec:model_FS},} then we describe the simulations conducted in \S\ref{sec:methods_FS}. The simulation results are compared with the model predictions in \S\ref{sec:results_FS}. This is followed by a discussion on our model's regime of validity and its practical implications in \S\ref{sec:discussion_FS}. Finally, the conclusions are drawn in \S\ref{sec:conclusion_FS}.

\begin{figure}
  \centerline{\includegraphics[width=0.8\linewidth]{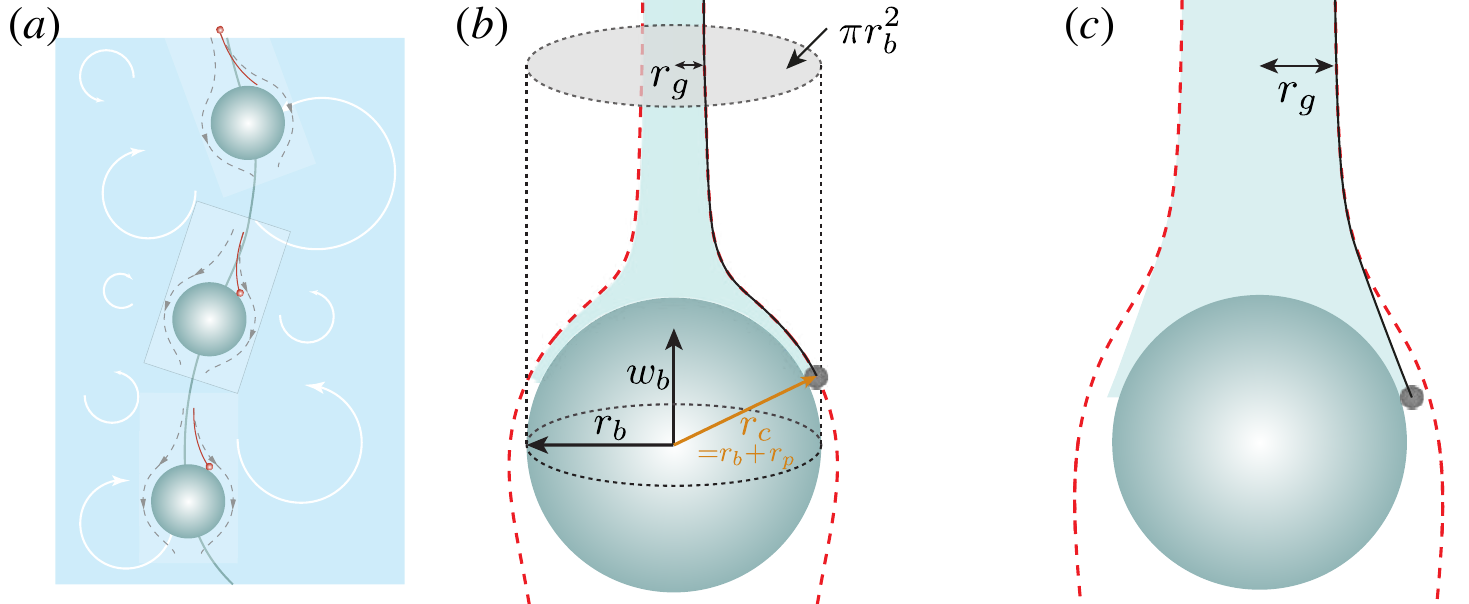}}
  \caption{({\textit{a}) Sketch of the 'frozen turbulence' collision model along the bubble trajectory where it collides with particles at multiple instances. The dashed line represents streamlines around the bubble and the red solid lines denote particle trajectories under the frozen turbulence approximation. The sizes of the bubble, the particles and the particle trajectories are exaggerated for visibility.}(\textit{b}) Sketch of a particle with $St_p' = 0$ travelling along the grazing trajectory (black solid line) to collide with a bubble in still fluid. The red dashed line is the corresponding streamline. The collision cylinder used to define the collision efficiency is also shown. (\textit{c}) Sketch of a bubble--particle collision with finite $St_p'$ where the grazing trajectory deviates from the corresponding streamline.}
  \label{fig:collisionsketch}
\end{figure}

\section{Modelling approach}\label{sec:model_FS}
\subsection{General concept}
We start with a brief description of the collision framework proposed in \citet{jiang_how_2024}. 
To obtain the collision kernel, the bubble trajectory is decomposed into tiny fragments, for which the bubble velocity can be considered constant. During the time it takes for the bubble to travel along such a fragment, a `frozen turbulence' approximation is employed (i.e. the flow field close to the bubble is considered to be steady and uniform), such that the instantaneous process is equivalent to the deterministic still fluid case with a bubble slip velocity corresponding to the instantaneous value. {This approximation therefore focuses on the turbulence effects on bubbles instead of particles since for finite-size bubbles, the flow near the bubble deviates from homogeneous isotropic turbulence such that turbulence effects on particles may be less significant. Figure \ref{fig:collisionsketch}(a) sketches the frozen turbulence approximation for an example bubble trajectory, where the streamlines and particle trajectories around the bubble positions at three instants are shown.} At every instant, the collision kernel can be evaluated by the collision efficiency with the corresponding conditions (see \citet{jiang_how_2024} for details).
Integrating this over the entire bubble trajectory results in the overall collision kernel 
\begin{equation}    \label{eq:collKFromEc}
    \Gamma = \pi r_b^2 \int_{0}^{\infty} E_c(Re_b,St_p')w_bf(w_b)\mathrm{d}w_b,
\end{equation}
where $f(\cdot)$ is the probability density function (p.d.f.) and $E_c$ represents the collision efficiency in a still fluid as a function of the bubble Reynolds number ($Re_b = 2r_b w_b / \nu$) and the particle Stokes number ($St_p' = \tau_p w_b / 2r_b$), both defined with respect to the bubble slip velocity $w_b$. 
As illustrated in figure \ref{fig:collisionsketch}(b), only particles that are initially (i.e. in the undisturbed flow far from the bubble) located within a distance $r_g$ from the centreline of the cylinder swept by the bubble eventually collide with the bubble ($r_g$ is hence known as the `grazing radius'). 
The collision efficiency $E_c$ can therefore be expressed as the ratio between the area of a circle with radius $r_g$ to that of a circle with radius $r_b$ \citep{schulze_hydrodynamics_1989,dai_inertial_1998}:
\begin{equation}
    E_c = \frac{\text{number of colliding particles per unit length}}{\text{number of particles swept by the bubble per unit length}}= \frac{n_p\pi r_g^2}{n_p\pi r_b^2} = \bigg(\frac{r_g}{r_b}\bigg)^2.
\end{equation}
In order to turn Eq.(\ref{eq:collKFromEc}) into a predictive model, parametrisations for both $f(w_b)$ and $E_c$ must be established. These will be presented in the following two sections, respectively.

\subsection{Parametrisation of the collision efficiency}
In general, $E_c$ is determined by five nondimensional parameters \citep{sarrot_determination_2005,huang_effect_2012,schuch_uber_1978,schulze_hydrodynamics_1989}: 
\begin{equation}    \label{eq:EcStillFluidParam}
    Re_b,\, \frac{r_b}{r_p},\, St_p',\, \frac{v_s}{w_b},\, \theta_{cap},
\end{equation}
which are the bubble Reynolds number, the bubble-particle size ratio, the particle Stokes number based on bubble slip velocity, the particle settling speed ratio, and the bubble contamination angle, respectively. Additionally, $v_s$ is the particle settling speed. 
In the discussion below, we restrict ourselves to the case of fully contaminated bubbles ($\theta_{cap} = \pi$), which is reasonable for bubbles in flotation \citep{nguyen_colloidal_2004,huang_effect_2012}.

In the $St_p' \to 0$ limit particles perfectly follow the fluid flow. A collision then occurs if the streamline carrying the particle comes within the collision distance $r_c$. This is termed the `interception mechanism' \citep{schulze_hydrodynamics_1989} and \citet{sarrot_determination_2005} showed that 
\begin{equation}    \label{eq:Ec_interception}
    E_c^{(i)} = \bigg(\frac{r_g}{r_b}\bigg)^2 = \frac{3}{2}\bigg(\frac{r_p}{r_b}\bigg)^2 \bigg(1 + \frac{Re_b^{2/3}}{5}\bigg).
\end{equation}
As $St_p'$ increases, the particle trajectories are less affected by the locally distorted flow field, which can cause more of the incident particles to collide with the front half of the bubble  as shown in figure \ref{fig:collisionsketch}(\textit{b}), thus increasing the collision efficiency \citep{huang_effect_2012}. This `inertial mechanism' can be expressed as \citep{schuch_uber_1978,schulze_hydrodynamics_1989}
\begin{center}
\begin{equation}    \label{eq:Ec_inertial}
    E_c^{(in)} = \bigg(1 + \frac{r_p}{r_b}\bigg)^2\bigg(\frac{St_p'}{St_p' + a}\bigg)^b,
\end{equation}
\end{center}
where $a$ and $b$ are fitting parameters, which were determined through additional simulations (see \ref{sec::schulzeFittingParameters}). 
Furthermore, gravitational settling of heavy particles also affects the collision efficiency. To capture this effect, \citet{weber_interceptional_1983} incorporated the particle settling velocity additively into the local fluid velocity at the collision position and integrated the resulting particle flux over the collision angle up to a critical value $\theta_c$, thereby deriving the gravitational collision efficiency:
\begin{equation}    \label{eq:Ec_inertial}
    E_c^{(g)} = -\left(1+\frac{r_p}{r_b}\right)^2\frac{v_s}{w_b}\sin^2\theta_c,
\end{equation}
where we adopt the expression of $\theta_c$ from \citet{kostoglou_generalized_2020} (also provided in \ref{sec::kostoglouExpressions}). Note that since particle settling corresponds to negative values of $v_s$, the resulting gravitational contribution to the collision kernel is positive in this case. 
\citet{schulze_hydrodynamics_1989} proposed to model the overall collision efficiency as a superposition of the interception, gravitational, and inertial contributions, leading to 
\begin{equation}    \label{eq:Ec_total}
    E_c = E_c^{(i)} {+E_c^{(g)}}+ E_c^{(in)}\bigg(1 - \frac{E_c^{(i)}}{(1 + r_p/r_b)^2}\bigg).
\end{equation}
Here, the term in brackets ensures that $E_c$ scales as $(1+r_p/r_b)^2$ in the high $St_p'$ limit. 
We note that the inertial and gravitational contributions are treated as additive, though this assumption is not strictly valid in every case. It is applied in this expression with the assumption that either $E_c^{(i)}+E_c^{(g)}$ or $E^{(in)}$ is dominant \citep{schulze_hydrodynamics_1989}. Nevertheless, this formulation represents, to the best of our knowledge, the most physically consistent and predictive expression currently available for describing the combined influence of particle inertia and gravity. 
Note that the particle gravitational settling effect is not included in the work of \citet{jiang_how_2024}. As the particle settling may have a role when the particle Stokes number is large, we here extend the model of \citet{jiang_how_2024} by including $E_c^{(g)}$.

\subsection{Modelling the bubble slip velocity probability density function}
The final missing ingredient to turn the approach outlined in Eq.(\ref{eq:collKFromEc}) into a predictive model is a suitable model that predicts the p.d.f. of the bubble slip speed $f(w_b)$ using flow and bubble parameters as inputs. For this, we start with considering the slip velocity vector $\mathbf{w}_b$ with normally distributed velocity components $w_{b,i}$ ($i=x,y,z$ indicates the direction), i.e.
\begin{equation}    \label{eq:pdfSlipVelNormal}
    f(w_{b,i}) = \frac{1}{\sqrt{2\pi \sigma^2_{i}}}\exp{\bigg( -\frac{(w_{b,i} - \langle w_{b,i} \rangle)^2}{2\sigma^2_{i}}\bigg)},
\end{equation}
where $\langle \cdot \rangle$ denotes averaging and $\sigma_{i}$ is the standard deviation of the bubble slip velocity component. This approach has also been taken by \citet{berk_analytical_2024}. For the horizontal components $\langle w_{b,x} \rangle = \langle w_{b,y} \rangle = 0$ as buoyancy acts only along the vertical ($z$) direction and hereafter we take $\langle w_{b,z} \rangle = \langle w_{b} \rangle$ for simplicity. We note that $f(w_{b,i})$ is sometimes better described by a stretched exponential function \citep{masuk_simultaneous_2021}. Nonetheless, as the main difference lies in the tails of the distributions, the impact on the predicted collision kernel $\Gamma$ is minimal \citep{bragg_new_2014,saffman_collision_1956}. Additionally, fitting parameters have to be introduced if a stretched exponential function is employed. We therefore choose to model $f(w_{b,i})$ with a Gaussian distribution. Assuming that the components are independent of each other and $\sigma_{x} = \sigma_{y} = \sigma_{z} = \sigma_i$ \citep{berk_analytical_2024}, we integrate the joint p.d.f. over a spherical shell $S$ with a radius of $\sqrt{w^2_{b,x} + w^2_{b,y} + w^2_{b,z}}$ to obtain
\begin{eqnarray}    \label{eq:pdfSlipVel}
    f(w_b) &=& \oiint_S f(w_{b,x})f(w_{b,y})f(w_{b,z}) \mathrm{d}S\nonumber\\
    &=& \frac{\sqrt{2}w_b}{\sqrt{\pi} \sigma \langle w_{b} \rangle}\exp{\bigg(-\frac{w_b^2 + \langle w_b \rangle^2}{2\sigma^2} \bigg)}\sinh{\bigg(\frac{w_b \langle w_b \rangle}{\sigma^2}\bigg)}.
\end{eqnarray}
We note that mathematically evaluating Eq. (\ref{eq:pdfSlipVel}) in the limiting of large $\langle w_b\rangle$ leads to undefined values, since $\lim_{x\to\infty} \sinh{(x)}/x \to \infty$ and $\lim_{x\to\infty}\exp{(-x^2)}\to 0$. This limiting case corresponds to a gravity-dominated regime in which the slip velocity vector is oriented nearly vertically and the velocity fluctuation is negligible. Such a scenario does not occur in the turbulent flow conditions we consider. Nevertheless, when the mean slip velocity is relatively large with respect to the fluctuations, the distribution $f(w_b)$ converges to a normal distribution centred at $\langle w_b\rangle$ (see Appendix C). 
Therefore, we use
\begin{align}   \label{eq:pdfSlipVelComplete}
    f(w_b) &=
		\begin{cases}
			\, \frac{\sqrt{2}w_b}{\sqrt{\pi} \sigma \langle w_{b} \rangle}\exp{\bigg(-\frac{w_b^2 + \langle w_b \rangle^2}{2\sigma^2} \bigg)}\sinh{\bigg(\frac{w_b \langle w_b \rangle}{\sigma^2}\bigg)} &\text{if $\langle w_b \rangle/\sigma \leq 16$,} \\
			\, \frac{1}{\sqrt{2\pi}\sigma}\exp\bigg(-\frac{(w_b - \langle w_b \rangle)^2}{2\sigma^2}\bigg) &\text{otherwise.}
		\end{cases}
\end{align}

Based on (\ref{eq:pdfSlipVelComplete}), $f(w_b)$ can be constructed if the mean vertical bubble slip velocity and the standard deviation of the slip velocity components are determined. The models for these quantities in literature are all based on the point-bubble approximation \citep{maxey_equation_1983,gatignol_faxen_1983}, which means that the bubble dynamics are governed by
\begin{eqnarray}\label{eq:PointParticleEoM_GFS}
\frac{4}{3}\pi r_b^3\rho_b\frac{\mathrm{d}\boldsymbol{v_b}}{\mathrm{d}t} & = & 6\pi\mu r_bf_b(\boldsymbol{u}-\boldsymbol{v_b}) + \frac{4}{3}\pi r_b^3\rho_f\frac{\mathrm{D}\boldsymbol{u}}{\mathrm{D}t} + \frac{2}{3}\pi r_b^3\rho_f\bigg(\frac{\mathrm{D}\boldsymbol{u}}{\mathrm{D}t} - \frac{\mathrm{d}\boldsymbol{v_b}}{\mathrm{d}t}\bigg)\nonumber\\ &&
+ \frac{4}{3}\pi r_b^3(\rho_f - \rho_b)\mathfrak{g}\boldsymbol{e}_z,
\end{eqnarray}
where $\boldsymbol{v_b}$ is the bubble velocity, $\boldsymbol{u}$ is the fluid velocity, $t$ is the time, $\mu=\nu\rho_f$ is the absolute viscosity, $f_b$ is the correction factor that accounts for nonlinear drag and is a function of $Re_b$, and $\boldsymbol{e}_z$ is the unit vector pointing vertically upwards. The terms on the right hand side of (\ref{eq:PointParticleEoM_GFS}) are the drag force, pressure gradient force, added mass force, and buoyancy, respectively. (\ref{eq:PointParticleEoM_GFS}) does not include the history force and the lift force as these are neglected in many models to improve tractability. We note that the lift force can play a significant role on the bubble trajectory \citep{spelt_motion_1997}. Nonetheless, for predicting the magnitude of the slip velocity, \citet{berk_analytical_2024} showed that neglecting the lift force does not significantly affect the result. Hence we do not consider the lift and history forces in our model.

\subsubsection{Mean vertical slip velocity}    \label{sec:model_meanVerticalSlip}
To obtain the mean vertical slip velocity, we assume that it is approximately equal to the bubble rise velocity. This is a strong assumption since it implies that the fluid sampled by the bubbles has a negligible average vertical velocity, even though bubbles are known to cluster in certain flow regions in turbulence \citep{calzavarini_quantifying_2008-1,calzavarini_dimensionality_2008}. This assumption will be justified in \S\ref{sec:results_FS}. \citet{ruth_effect_2021} investigated the bubble rise velocity in turbulence by introducing a Reynolds decomposition of $\boldsymbol{v_b} = \langle v_{b,z} \rangle \boldsymbol{e}_z + \boldsymbol{\Tilde{v}_b}$ and in the absence of a mean flow $\boldsymbol{u} = \boldsymbol{\Tilde{u}}$ into (\ref{eq:PointParticleEoM_GFS}), where $\Tilde{\cdot}$ represents fluctuations. For small $1/Fr$ (i.e. weak gravity), they found that turbulence reduces the bubble rise velocity through nonlinear drag, which couples the horizontal slip velocity components with the vertical component as $Re_b$ depends on the magnitude of the slip velocity. As a result, $\langle v_{b,z} \rangle/v_q = 0.37/Fr_L$, where $v_q$ is the bubble rise velocity in quiescent fluid, $Fr_L = u'/\sqrt{2\mathfrak{g} r_b}$ is a large-scale Froude number, and 0.37 is obtained by fitting. For the large $1/Fr$ limit, they also found that turbulence reduces the bubble rise velocity. To model this, they considered $|\boldsymbol{v_b} - \boldsymbol{u}| \approx v_q$ as well as $\langle |\boldsymbol{\Tilde{v}_b} - \boldsymbol{\Tilde{u}}|\rangle \propto u'$ which was observed in their simulations, and expanded $|\boldsymbol{v_b} - \boldsymbol{u}|$ around $v_q$. This resulted in $\langle v_{b,z} \rangle/v_q = (1-\chi Fr_L^2)$, where $\chi = \langle |\boldsymbol{\Tilde{v}_b} - \boldsymbol{\Tilde{u}}|\rangle/(4u')$ is a proportionality constant. We adopt these parametrisations here and model the mean vertical slip velocity with
\begin{equation}    \label{eq:meanSlip}
\langle w_b \rangle = 
\left\{
\begin{aligned}
v_q (1- \chi Fr_L^2)     \quad&\text{for $1/Fr_L \geq 2.08$,}\\
0.37 \frac{v_q}{Fr_L} \quad&\text{otherwise,}\\
\end{aligned}
\right.
\end{equation}
where $v_q$ is determined through balancing the drag term with buoyancy in (\ref{eq:PointParticleEoM_GFS}) taking $f_{b} = 1 + 0.169(2r_bv_q/\nu)^{2/3}$, and $\chi = 1$ in accordance with \citet{liu_direct_2024}.

\subsubsection{Standard deviation of the slip velocity}    \label{sec:stdSlipVel}
The other quantity that needs to be determined to construct $f(w_b)$ is the standard deviation of the bubble slip velocity $\sigma$. Several expressions for $\sigma$ have been proposed over the years. One of the earliest predictions draws analogy with gravitational settling \citep{liepe_untersuchungen_1976}. In this framework, gravitational acceleration is replaced by the centripetal acceleration $a_m$ of eddies with a size of $r_b$. The resultant apparent weight is then balanced by a drag force $\propto w_b^{3/2}$ (i.e. $f_b \propto \sqrt{w_b}$) and an assumption of $w_b\sim\sigma_i$ is applied. Considering eddies that lie inside the inertial subrange such that $a_m = 2(\varepsilon r_b)^{2/3}/r_b$,
\begin{equation}    \label{eq:LMslip}
    \sigma_i^{(LM)} = \zeta \frac{\varepsilon^{4/9}r_b^{7/9}}{\nu^{1/3}}\bigg(\frac{\rho_f-\rho_b}{\rho_f}\bigg)^{2/3},
\end{equation}
where $\zeta$ is a constant whose value ranges from 0.57 to 0.83 in the literature \citep{liepe_untersuchungen_1976,schubert_turbulence-controlled_1999,nguyen_colloidal_2004}. Among these, \citet{kostoglou_critical_2020} argued that a value of 0.83 is correct which corresponds to $f_{bd} = (5/12)\sqrt{Re_b}$. We note that $\sigma_i^{(LM)}$ does not necessarily have to be motivated using gravitational settling. Instead, (\ref{eq:LMslip}) can be recovered in the no-gravity $St \ll 1$ limit of (\ref{eq:PointParticleEoM_GFS}) if the fluid acceleration is approximated by $a_m$, as discussed in \citet{chan_bubbleparticle_2023}. In this case, one still needs to take $w_b\sim\sigma_i$ and this assumption should not be taken for granted.

The model used in \citet{kostoglou_generalized_2020} does not employ this assumption and is based on the work about collisions between particles with different response times by \citet{yuu_collision_1984}, which has been extended to bubble--particle collisions in \citet{ngo-cong_isotropic_2018}. For the bubble slip velocity, \citet{kostoglou_generalized_2020} considers the special case where one of the particles is a tracer. As the buoyancy term is constant over time, it is not expected to directly affect $\sigma_i$ and is therefore omitted from this analysis. In this model, the standard deviation is determined through
\begin{equation}
    \sigma_i^2 = \langle(\Tilde{v}_{b,i} - u_i)^2\rangle = \langle \Tilde{v}_{b,i}^2\rangle + u_i'^2 - 2\langle \Tilde{v}_{b,i} u_i\rangle.
\end{equation}
The bubble velocity variance $\langle \Tilde{v}_{b,i}^2\rangle$ is obtained by integrating the bubble energy spectrum $E_b(\omega)$, which is related to the fluid energy spectrum at the particle location $E(\omega)$ such that
\begin{equation}    \label{eq:KosBubbleRMSBasic}
    \langle \Tilde{v}_{b,i}^2\rangle = \int_{0}^{\infty} E_b(\omega) \mathrm{d}\omega = \int_{0}^{\infty} H^2(\omega)E(\omega) \mathrm{d}\omega,
\end{equation}
where the response function {of the bubble to turbulent fluctuations is given by}
\begin{equation}    \label{eq:bubResponseFunc}
    H^2(\omega) = \frac{1 + (\beta\omega\tau_p/f_b)^2}{1 + (\omega\tau_p/f_b)^2}
\end{equation}
and
\begin{equation}    \label{eq:kosEnergySpec}
    E(\omega) = E^{(Kos)}(\omega) = u'^2\frac{T_L}{1 + (\omega T_L)^2}.
\end{equation}
$\beta = 3\rho_f/(2\rho_b + \rho_f)$, and $T_L$ is the Lagrangian time scale. The procedure for evaluating $\langle \Tilde{v}_{b,i} u\rangle$ is the same as (\ref{eq:KosBubbleRMSBasic}) apart from a slightly different $H^2(\omega)$. For both terms, $H^2(\omega)$ is obtained by expanding $\boldsymbol{v_b}$ and $\boldsymbol{u}$ with a Fourier integral in the point-bubble equation. Crucially, $\mathrm{d}/\mathrm{d}t = \mathrm{D}/\mathrm{D}t$ was taken in the point-bubble equation (\ref{eq:PointParticleEoM_GFS}), which implies that the resulting expression
\begin{equation}    \label{eq:KosBubbleRMS}
    \sigma_i = \frac{2u'}{\sqrt{1 + T_L f_b/\tau_b}}
\end{equation}
is strictly true only when $St_b \ll 1$. Substituting $T_L/\tau_\eta = 0.181Re_\lambda$ and $f_b = 1 + 0.169(2r_b\sigma_i/\nu)^{2/3}$ \citep{kostoglou_generalized_2020,nguyen_colloidal_2004}, (\ref{eq:KosBubbleRMS}) becomes
\begin{equation}    \label{eq:KosBubbleRMSwithfb}
    \sigma_i^{(Kos)} = 2u_i'\bigg(1 + 8.897\frac{u_i'^2\sqrt{\nu \sigma_i^{(Kos)}}}{\varepsilon r_b^{3/2}}\bigg)^{-1/2}.
\end{equation}
Note that as the expression of $f_b$ used tacitly implies $Re_b = 2r_b\sigma_i/\nu$, so (\ref{eq:KosBubbleRMSwithfb}) should only apply for $1/Fr = 0$.

Most recently, \citet{berk_analytical_2024} used an alternative approach to derive an expression for $\sigma_i$ by considering the variance of (\ref{eq:PointParticleEoM_GFS}):
\begin{equation}    \label{eq:BCsigmaBegin}
\sigma_i^2 = \frac{\tau_b^2}{f_b^2}\bigg\langle\bigg(\frac{\mathrm{d}v_{b,i}}{\mathrm{d}t}\bigg)^2\bigg\rangle + \frac{\tau_b^2\beta^2}{f_b^2}\bigg\langle\bigg( \frac{\mathrm{D}u_i}{\mathrm{D}t}\bigg)^2\bigg\rangle - 2\frac{\tau_b^2\beta}{f_b^2}\bigg\langle\frac{\mathrm{d}v_{b,i}}{\mathrm{d}t}\frac{\mathrm{D}u_i}{\mathrm{D}t}\bigg\rangle,
\end{equation}
where using (\ref{eq:PointParticleEoM_GFS}) the last term is expressed as
\begin{equation}    \label{eq:BCcrossTerm}
    \bigg\langle\frac{\mathrm{d}v_{b,i}}{\mathrm{d}t}\frac{\mathrm{D}u_i}{\mathrm{D}t}\bigg\rangle = \frac{f_b}{\tau_b}\bigg\langle w_{b,i}\frac{\mathrm{D}u_i}{\mathrm{D}t}\bigg\rangle + \beta\bigg\langle\bigg( \frac{\mathrm{D}u_i}{\mathrm{D}t}\bigg)^2\bigg\rangle.
\end{equation}
The bubble acceleration variance $\langle(\mathrm{d}v_{b,i}/\mathrm{d}t)^2 \rangle$ and fluid acceleration variance $\langle (\mathrm{D}u_i/\mathrm{D}t)^2 \rangle$ are then obtained by integrating the particle ($\omega^2 H^2(\omega) E(\omega)$) and fluid acceleration spectra ($\omega^2 E(\omega)$), respectively, where $H^2(\omega)$ is given by (\ref{eq:bubResponseFunc}) and $E(\omega)$ is a model energy spectrum with two time scales \citep{sawford_reynolds_1991}
\begin{equation}    \label{eq:BCEnergySpec}
    E(\omega) = E^{(BC)}(\omega) = u'^2\frac{T_L + T_2}{(1 + (\omega T_L)^2)(1 + (\omega T_2)^2)}.
\end{equation}
Here, $T_L/\tau_\eta = 2(Re_\lambda + 32)/(\sqrt{15}C_0)$, $T_2 = C_0/2(a_0)$, $C_0 = 7$, and $a_0 = 5/(1 + 110/Re_\lambda)$. The short time scale $T_2$ allows (\ref{eq:BCEnergySpec}) to faithfully capture high-frequency behaviour by constraining the slope of the corresponding velocity correlation function to zero at zero time lag. Note that we take $C_0 = 7$ instead of the expressions listed in \citet{berk_analytical_2024} in view of the better agreement with the values of $T_L$ reported in \citet{zaichik_two_2003}.
On the other hand, the covariance term $\langle w_{b,i}(\mathrm{D}u_i/\mathrm{D}t)\rangle$ in (\ref{eq:BCcrossTerm}) is evaluated by assuming $\mathrm{d}\boldsymbol{v_b}/\mathrm{d}t = \mathrm{D}\boldsymbol{u}/\mathrm{D}t$ in (\ref{eq:PointParticleEoM_GFS}) such that
\begin{equation}    \label{eq:BCcovariance}
    \bigg\langle w_b \frac{\mathrm{D}u_i}{\mathrm{D}t}\bigg\rangle = \frac{\sigma_i^2f_b}{\tau_b(1 - \beta)}.
\end{equation}
Finally, using (\ref{eq:BCcrossTerm}) -- (\ref{eq:BCcovariance}) in (\ref{eq:BCsigmaBegin}) yields
\begin{equation}    \label{eq:BCBubbleRMS}
    \sigma_i^{(BC)} = u'(\beta - 1)\frac{\tau_b/f_{b}}{\sqrt{(T_L + \tau_b/f_b)(T_2 + \tau_b/f_b)}},
\end{equation}
where here we choose $f_{b} = 1 + 0.169(2r_b\langle w_b\rangle/\nu)^{2/3}$ \citep{nguyen_colloidal_2004} with $\langle w_b \rangle$ defined by (\ref{eq:meanSlip}). Note that other expressions of $f_b$ may be used as desired.

Throughout the derivation of (\ref{eq:BCBubbleRMS}), particle accelerations are always related to the fluid accelerations using either (\ref{eq:bubResponseFunc}) or (\ref{eq:BCcovariance}), both of which are valid for the $St\ll 1$ limit. Hence, this technically implies that $\sigma_i^{(BC)}$ is only applicable for $St\ll 1$. In fact, (\ref{eq:BCBubbleRMS}) reduces to (\ref{eq:KosBubbleRMS}) when $\beta = 3$ (corresponding to $\rho_b = 0$) and $T_2 = 0$. Despite their similarities, there are two key differences between $\sigma_i^{(Kos)}$ and $\sigma_i^{(BC)}$. First, $f_b$ that enters $\sigma_i^{(Kos)}$ does not include the mean settling velocity. Second, \citet{kostoglou_generalized_2020} employs an expression for $E(\omega)$ with a single time scale. As will be shown in \S\ref{sec:results_FS}, (\ref{eq:BCBubbleRMS}) gives a better prediction of $\sigma_i$ and will be the one we select to model the bubble slip velocity p.d.f. as displayed in (\ref{eq:pdfSlipVelComplete}).

\section{Numerical methods}\label{sec:methods_FS}
\subsection{Equations and solvers}  \label{sec:methods_eq_FS}
In order to assess the quality of our model predictions, we perform direct numerical simulations of interface-resolved bubbles and point-particles in homogeneous isotropic turbulence.
The fluid flow is governed by the incompressible Navier--Stokes equations and the continuity equation, which read:
\begin{eqnarray}
		\frac{\partial\textbf{u}}{\partial t}  + \textbf{u}\cdot {\nabla}   \textbf{u} &=& -   \frac{1}{\rho_f}{\nabla} p + \nu\ \nabla^2   \textbf{u} + \textbf{f} + \textbf{f}_b, \label{eq:N-S}\\
		{\nabla}  \cdot  \textbf{u} &=& 0, \label{eq:div}
\end{eqnarray}
where $\mathbf{u}$ is the fluid velocity, $p$ denotes the pressure, and $\mathbf{f_b}$ is the force exerted by the bubble on the fluid which will be discussed later in this section. 
$\mathbf{f}$ is an external random large-scale volume force that sustains the background turbulence \citep{Perlekar2012}. 
This volume force term $\mathbf{f}$ is statistically homogeneous and isotropic with a time-independent global energy input, such that the resulting turbulence is statistically stationary, homogeneous and isotropic. To obtain the flow field, (\ref{eq:N-S}) and (\ref{eq:div}) are solved on a uniform Eulerian grid using the lattice Boltzmann code developed in \citet{Calzavarini2019} and \citet{Jiang2022}.

To simulate the bubbles, we consider them as two-way coupled buoyant spheres with zero slip on the surface. This is a reasonable assumption for bubbles in flotation processes since a significant amount of surfactants is usually present in the liquid, such that the bubble surface can be considered fully contaminated and the no-slip condition can be applied \citep{nguyen_colloidal_2004, huang_effect_2012}. We note that surfactants can reduce the surface tension and promote bubble deformation. As an indicator of the deformation, we calculate the Weber number, $We=2r_b\rho_f \langle v_{b,z} \rangle^2/\gamma$, based on the surface tension $\gamma$ of water and the measured bubble rise velocity in turbulence $\langle v_{b,z} \rangle$, and find that it is $\mathrm{\textit{O}}(0.1)$ for our simulations meaning bubble deformation is small ($We$ is even smaller if the turbulent fluctuation is considered instead of $\langle v_{b,z} \rangle$). We furthermore focus on bubbles with an average $Re_b \lesssim 200$, which according to \citet{clift_bubbles_2005} are spherical in shape. Consequently, we model the bubbles as rigid spheres whose translational and rotational dynamics are governed by the Newton--Euler equations
\begin{eqnarray}
		m_b \frac{{\rm d} \mathbf{v}_b}{{\rm d}t} &=& \oiint_{S_b} \bm{\sigma} \cdot \mathbf{n}\ {\rm d}S - V_b(\rho_b-\rho_f)\mathfrak{g}\mathbf{e}_z, \label{eq:Newton-Euler1}\\
		\frac{{\rm d}  \bm{\mathbf{I_b}} \mathbf{\Omega}_b}{{\rm d}t} &=& \oiint_{S_b} (\mathbf{x}-\mathbf{x}_b) \times  (\bm{\sigma} \cdot \textbf{n})\ {\rm d}S. \label{eq:Newton-Euler}
\end{eqnarray}
Here, $\mathbf{\Omega}_b$ is the angular velocity of the bubble at position $\mathbf{x}_b$ with a volume $V_b$ and a surface $S_b$, $m_b =\rho_b V_b$ is the bubble mass, $\bm{\mathbf{I_b}}$ is its moment of inertia tensor, 
$\textbf{n}$ is a unit vector pointing outward normal to the bubble surface, and $\mathbf{x}$ is the position vector.
The fluid stress tensor $\bm{\sigma} = -p\boldsymbol{I}+\mu(\nabla \boldsymbol{u} + \nabla\boldsymbol{u}^T)$, where $\boldsymbol{I}$ is the identity tensor, is solved by the immersed boundary method (IBM) \citep{verzicco_immersed_2023,Uhlmann2005}. This method imposes a volume force $\mathbf{f}^{ibm}$ on the bubble--liquid interface that couples the bubble with the liquid, and vice versa.
To compute the bubble motions and determine $\mathbf{f}^{ibm}$, we use the semi-implicit method described in \citet{Tschisgale2017}, instead of the conventional explicit method which is numerically unstable when the bubble density is low \citep{Uhlmann2005}.

A potential artefact of taking $\mathbf{f}_b = \mathbf{f}^{ibm}$ in our present case is the formation of a mean upward flow driven by the rising motion of the bubbles. 
To avoid such a situation, we require the net force exerted on the fluid by the bubble to be zero \citep{Hoefler2000, chouippe_forcing_2015}.
This is achieved by subtracting from $\mathbf{f}^{ibm}$ its spatial average
\begin{eqnarray}
	\langle \mathbf{f}^{ibm}\rangle_\Pi(t) = \frac{1}{||\Pi||}\int_\Pi \mathbf{f}^{ibm}(\mathbf{x},~t) \rm{d}\mathbf{x} \label{eq:mean_ibm_force}
\end{eqnarray}
 at every time step, where $\Pi$ denotes the entire computational domain, $\langle \cdots\rangle_\Pi$ indicates spatial averaging over $\Pi$, and $||\Pi||$ is the volume of $\Pi$. Hence, the force exerted by the bubble on the fluid is 
\begin{eqnarray}
	\mathbf{f}_b(\mathbf{x},t)=\mathbf{f}^{ibm}(\mathbf{x},t)-\langle \mathbf{f}^{ibm}\rangle_\Pi(t). \label{eq:compens_ibm}
\end{eqnarray}

In contrast to the bubbles, the size of the mineral particles in the flotation processes are comparable to $\eta$, meaning they can be modelled as point-like particles. Here, we initially consider one-way coupled non-settling particles that can overlap with each other and experience the drag, pressure gradient, and added mass forces, which is consistent with the modelling assumptions in \citet{jiang_how_2024}. Additionally, complementary simulations incorporating particle gravity are performed to examine the effect of particle settling on the collision kernel. The particle dynamics is therefore governed by (\ref{eq:PointParticleEoM_GFS}), except all bubble quantities are replaced by those of the particles, and {the drag coefficient is} $f_p=1+0.15Re_p^{0.687}$.
The particle equation is solved using a finite-difference scheme where the spatial derivatives are discretised using a second-order central finite-difference method and time marching is performed with a second-order Adams--Bashforth scheme using the same time step size as the fluid solver. Since the terms involving $\mathbf{u}$ need to be evaluated at particle positions, their values are interpolated from the Eulerian grid using a tri-linear scheme.

\subsection{Simulation parameters}  \label{sec:methods_sim_FS}
The equations described in \S \ref{sec:methods_eq_FS} are solved to simulate bubbles with $\rho_b/\rho_f = 1/1000$, $St_b = 0.5 - 6.3$ and $1/Fr = 1-10$ in homogeneous isotropic turbulence with $Re_\lambda = 64$. Table \ref{tab:turbStat_FS} lists the parameters of the simulations and the turbulence statistics. We limit ourselves to low bubble volume fractions $\phi$ ($\leq 10$ bubbles) where bubbles rarely approach each other. Nonetheless, in case this happens, we employ a soft collision model with a coefficient of restitution of 0.8 to properly capture the lubrication force at small separations and prevent coalescence \citep{costa_collision_2015}. Our simulations are performed in a triply periodic domain. 
Hence, in principle it is possible for a rising bubble to encounter its own wake, which would affect the statistics. In the wake, the velocity defect $\delta u_b$ decays initially with the downstream distance $z$ as $z^{-1}$ until it reaches the same order of magnitude as the turbulent velocity fluctuations at $z_0$, beyond which it decays as $z^{-2}$ \citep{Legendre2006pof}. 
We take the velocity defect at the bubble position $\delta u_b(0)=\langle v_{b,z} \rangle$, define $z_0$ by $\delta u_b(z_0) = u'$, and estimate the velocity defect one domain length downstream of the bubble $\delta u_b(L_z) = 0.01 u' \ll u'$, which is negligibly small. Additionally, we ran an extra simulation with $(St_b,1/Fr) = (6,10)$ in a longer domain and find that the statistics do not change significantly. These considerations suggest that the bubble's motion is not significantly affected by its own wake.
To ensure sufficient spatial resolution of the boundary layer on the bubbles, we estimate the thickness of the boundary layer at the front stagnation point as $\delta /(2r_b)=1.13/Re_b^{1/2}$ following \citet{johnson_flow_1999}, and resolve this by at least 7 grid points. A grid convergence study (see \ref{sec::GridConvergence}) confirmed that this is sufficient for the present cases. The bubbles are initialised on a primitive cubic lattice with the fluid velocity at their positions and allowed to adapt to the flow for 10 large eddy turnover times before their statistics are collected. The instantaneous fluid velocity at bubble position is determined by averaging the fluid velocity within a distance of $3r_b$ from the bubble, which for the case of a bubble subjected to a steady uniform flow recovers a value that reaches 90\% of the ground truth \citep{kidanemariam_direct_2013}.

\begin{table}
  \begin{center}
  \begin{tabular}{p{1cm}p{2cm}p{3cm}p{1cm}p{1cm}p{1cm}p{3cm}}
      \centering $St_b$ & $1/Fr$& $N_x\times N_y\times N_z$  & $r_b/\Delta x$ & $\eta/\Delta x$ & $\tau_\eta/\Delta t$  & $\phi$  \\[3pt]
      \hline
      \centering 0.5 & 1,2,4,10 & $512\times512\times768$ & 8.5 & 4 & 400 & $1.5\times10^{-4}$\\
      \centering 1 & 1,2,4,10 & $512\times512\times768$ & 12 & 4 & 400 & $4.3\times10^{-4}$\\
      \centering 2.8 & 1,2,4,10 & $768\times768\times1152$ & 30 & 6 & 1200 & $1.3\times10^{-3}$\\
      \centering 6.3 & 1,2 & $512\times512\times768$ & 30 & 4 & 800 & $4.5\times10^{-3}$\\
      \centering 6.3 & 4,10 & $768\times768\times1152$ & 45 & 6 & 1200 & $4.5\times10^{-3}$\\
  \end{tabular}
  \caption{Simulation parameters and the turbulence statistics for the different cases: the grid size for the whole simulation domain $N_x\times N_y\times N_z$, the grid spacing $\Delta x$, the time step size $\Delta t$, the Kolmogorov length scale $\eta$ and time scale $\tau_\eta$, the bubble volume fraction $\phi$. $Re_\lambda = 64$ for all cases.}
  \label{tab:turbStat_FS}
  \end{center}
\end{table}

For $St_b = 2.8$ and $6.3$, $2\times10^6$ point-particles with $r_p = r_b/30$ and various densities ($St_p\in [10^{-2}, 4]$) are injected randomly and homogeneously into the simulation domain. 
The accuracy of the particle dynamics near the bubble is ensured by requiring the Eulerian grid spacing to be smaller than the particle radius, so that the fluid velocity in the particle equation of motion is still correctly interpolated to the particle position even when the particle is close to the bubble. The addition of the particles allows us to determine the bubble--particle collision kernel using (\ref{eq:collKRate}) by measuring the collision rate directly, which we begin to do after $20\tau_\eta$. A collision is registered each time the distance between a particle and a bubble becomes smaller than the collision distance $r_c = r_p+r_b$. Note that at collision, the particle velocity relative to the bubble is small for low $St_p$ particles due to the no-slip boundary condition on the bubble surface. To prevent the particle from lingering around the bubble after collision, which would lead to multiple collisions with the same bubble, the particle is immediately reseeded at a random position that is at least $5r_b$ below the bubble after each collision.

\section{Results}\label{sec:results_FS}
\subsection{Mean vertical bubble slip velocity}
\begin{figure}
  \centerline{\includegraphics{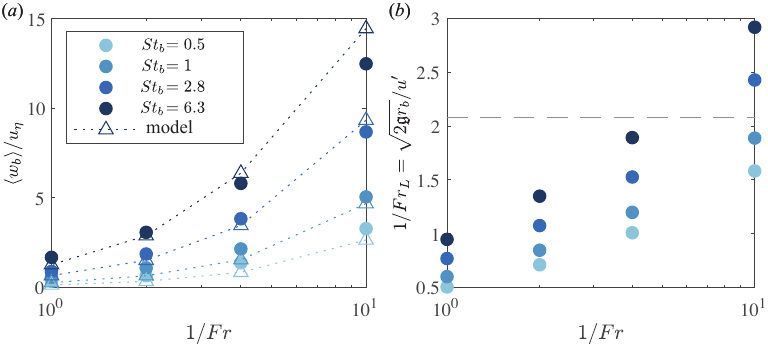}}
  \caption{(\textit{a}) The actual and predicted mean vertical bubble slip velocities and (\textit{b}) the inverse large-scale Froude number $1/Fr_L$ as a function of $1/Fr$. The dashed line in (\textit{b}) shows the critical $Fr_L$ corresponding to a change in the scaling of $\langle w_b \rangle$ in (\ref{eq:meanSlip}).}
\label{fig:meanSlipVel}
\end{figure}

We compare the predicted values of the mean vertical bubble slip velocity $\langle w_b \rangle$ with the actual values in figure \ref{fig:meanSlipVel}(\textit{a}). The data shows that $\langle w_b \rangle$ increases with $St_b$ and $1/Fr$, which is nicely captured by our model qualitatively and quantitatively. This justifies our use of (\ref{eq:meanSlip}) to model $\langle w_b \rangle$, even though (\ref{eq:meanSlip}) was developed for the bubble rise velocity instead of the slip velocity in turbulence. We additionally review the mechanism that reduces the bubble rise velocity in turbulence for the different cases by showing the values of $1/Fr_L$ in figure \ref{fig:meanSlipVel}(\textit{b}). For all the tested parameters except $(St_b,1/Fr) = (2.8,10)$ and $(6.3,10)$, $1/Fr_L$ is less than the transition value of 2.08 that separates the two scalings in (\ref{eq:meanSlip}). This suggests that for the vast majority of the simulated cases, nonlinear drag plays a dominant role in reducing the mean vertical rise velocity in turbulence. Since the transition $1/Fr_L = 2.08$ already corresponds to $\langle w_b \rangle/v_q = 0.77$ and $\langle w_b \rangle/v_q$ only reduces further as $1/Fr_L$ becomes smaller, using $v_q$ as a proxy for $\langle w_b \rangle$ (as is done in \citet{berk_analytical_2024}) would significantly overpredict the value of $\langle w_b \rangle$ for most of the tested cases.

\subsection{Standard deviation of the bubble slip velocity component}   \label{sec::results_StdDev}
\begin{figure}
  \centerline{\includegraphics{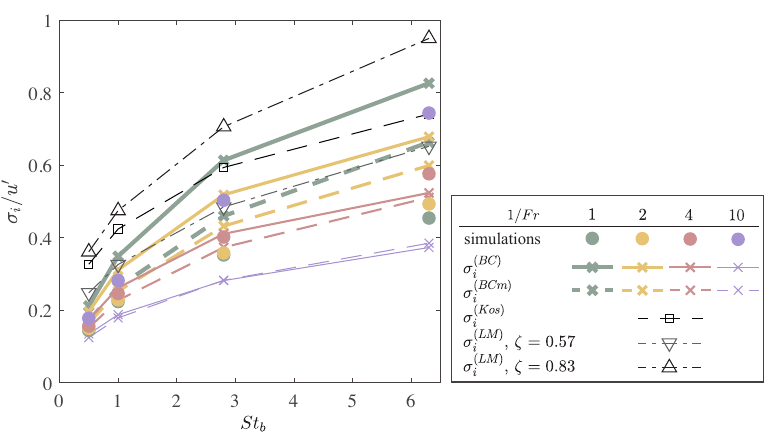}}
  \caption{Plot of the standard deviation of the bubble slip velocity as a function of $St_b$. $\sigma_i^{(BCm)}$ is the model by \citet{berk_analytical_2024} where the actual value of $f_b = 1 + 0.169\langle Re_b \rangle^{2/3}$ from simulations is used in (\ref{eq:BCBubbleRMS}). $\sigma_i^{(Kos)}$ and $\sigma_i^{(LM)}$ are independent of $1/Fr$.}
\label{fig:stdSlipVel}
\end{figure}

We next consider the other quantity needed to construct $f(w_b)$, which is the standard deviation of the slip velocity component $\sigma_i = \sqrt{(\sigma_x^2 + \sigma_y^2 + \sigma_z^2)/3}$. Figure \ref{fig:stdSlipVel} shows that $\sigma_i$ increases with both $St_b$ and $1/Fr$. As $St_b$ increases, the bubble is less responsive to higher frequency fluid fluctuation hence $\sigma_i$ increases. On the other hand, when $1/Fr$ increases, buoyancy becomes stronger and the bubbles rise faster, which means that they have less time to interact with the same eddy \citep{csanady_turbulent_1963}. This `crossing trajectory' effect causes $\sigma_i$ to increase. All of the models discussed in \S\ref{sec:stdSlipVel} qualitatively show the same trend in $St_b$ as the data, but none of them predict $\sigma_i$ increases with $1/Fr$. This is because these models do not account for the crossing trajectory effect, which reduces $T_L$ as $1/Fr$ increases \citep{csanady_turbulent_1963}. The only reason why $\sigma_i^{(BC)}$ (and $\sigma_i^{(BCm)}$ which will be introduced in the following paragraph) decreases with increasing $1/Fr$ is due to a larger $f_b$ resulting from a larger bubble rise velocity. Although it is possible to include expressions for $T_L$ that vary with $1/Fr$ \citep{berk_dynamics_2021}, it complicates the model as $\sigma_i$ would then be different for the vertical and horizontal components. We choose not to include this in our model since it is not of primary importance to the predicted collision kernel in the regime where our model is valid, as will be demonstrated later in this section.

To compare the models quantitatively, we consider the lowest $1/Fr$ case where gravity is weakest as $\sigma_i^{(LM)}$ and $\sigma_i^{(Kos)}$ do not account for the mean bubble rise velocity. Since the model by \citet{berk_analytical_2024} depends on the value of the drag correction $f_b$, we plot (\ref{eq:BCBubbleRMS}) using the measured value of $f_b$, i.e. $\sigma_i^{(BCm)}$, in addition to $\sigma_i^{(BC)}$, where as a first approximation $f_{b} = 1 + 0.169(2r_b\langle w_b\rangle/\nu)^{2/3}$ is taken by neglecting the slip velocity fluctuations. Among the models, the one by \citet{berk_analytical_2024} ($\sigma_i^{(BCm)}$) agrees the best with the data across the entire tested range of $St_b$, even though in principle it is valid only for $St_b \ll 1$; whereas the other models overpredict $\sigma_i$ especially when $St_b$ is small. For $\sigma_i^{(LM)}$, the discrepancy originates from its heuristic approach of balancing drag with buoyancy and the assumption that $\sigma_i \sim w_b$. As a result, it only gives an order-of-magnitude estimate of $\sigma_i$ especially when the correct value of $\zeta$ ($=0.83$) is considered. Meanwhile $\sigma^{(Kos)}$ overpredicts $\sigma_i$ because it uses a simpler energy spectrum $E(\omega)$ (see (\ref{eq:kosEnergySpec})) and neglects the mean bubble rise velocity in the drag correction $f_b$. In view of the above, we employ the prediction by \citet{berk_analytical_2024} in our model of the slip velocity p.d.f.. To keep the model simple, $\sigma^{(BC)}$ instead of $\sigma^{(BCm)}$ will be used. The difference between the two diminishes for large $1/Fr$ as $Re_b$ is increasingly dominated by the mean vertical slip velocity.

\begin{figure}
  \centerline{\includegraphics{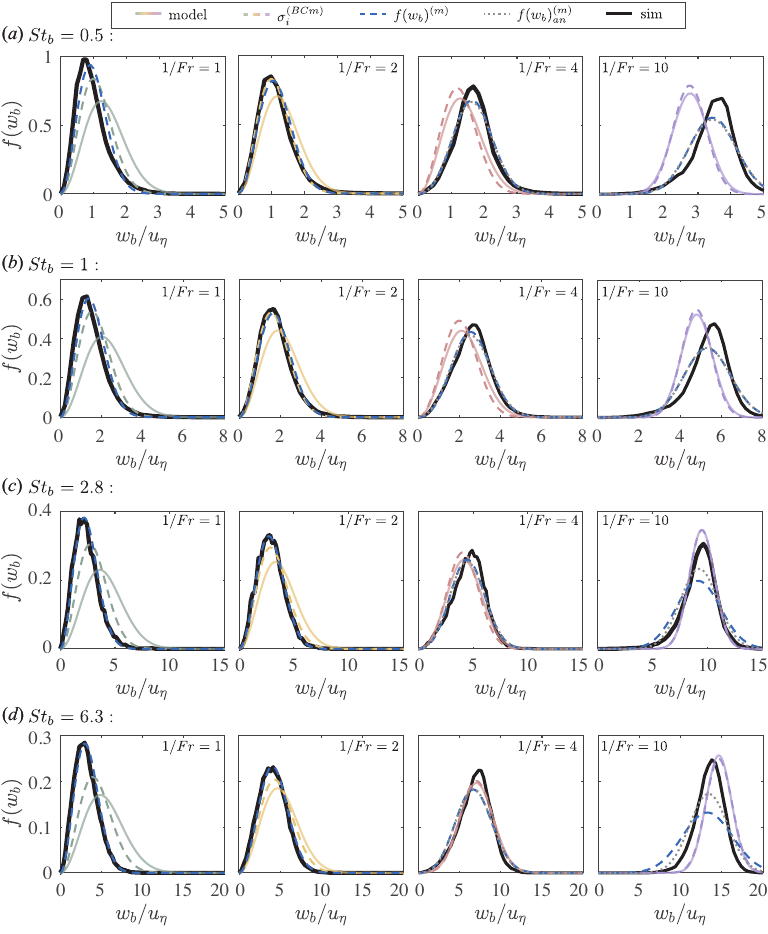}}
  \caption{Plot of the p.d.f. of the slip velocity magnitude at (\textit{a}) $St_b = 0.5$, (\textit{b}) $St_b = 1$, (\textit{c}) $St_b = 2.8$ and (\textit{d}) $St_b = 6.3$. The black line shows the actual p.d.f. from the simulations. The mean vertical slip velocities $\langle w_b \rangle$ used to obtain the model predictions (coloured solid lines) are given by (\ref{eq:meanSlip}). The dotted line shows the p.d.f. that is reconstructed from the measured $\langle w_b \rangle$ and $\sigma_i$.}
\label{fig:pdfSlipVel}
\end{figure}

\subsection{Bubble slip velocity probability density function}  \label{sec::results_slipVelPDF}
Having examined the mean and the standard deviations of the bubble slip velocity components, we focus on the slip velocity magnitude p.d.f. $f(w_b)$ in figure \ref{fig:pdfSlipVel}, where the {coloured} dashed lines show the p.d.f. reconstructed using the measured value of the bubble drag correction $f_b$ and will be discussed in \S\ref{sec:discussionValidity_FS}. The model prediction is generally good when $1/Fr \geq 4$ and the small shift in $f(w_b)$ is due to the slight difference between the actual $\langle w_b\rangle$ and the predicted value. This agreement is surprising given our model significantly underpredicts $\sigma_i$ when $1/Fr = 10$ as shown by figure \ref{fig:stdSlipVel}.

To investigate this, we test equation (\ref{eq:pdfSlipVel}) directly  by computing $f(w_b)$ obtained with the actual $\langle w_i \rangle$ and {$\sigma_i = \sqrt{(\sigma_x^2 + \sigma_y^2 + \sigma_z^2)/3}$}. The results{, $f(w_b)^{(m)}$,} are shown as {blue dashed} curves in figure \ref{fig:pdfSlipVel}. These curves agree excellently with the actual $f(w_b)$ when $1/Fr \leq 2$, but they are wider than the actual $f(w_b)$ at larger $1/Fr$. {To pinpoint the source for this discrepancy, we relax the assumption that $\sigma_x = \sigma_y = \sigma_z$ and reconstruct $f(w_b)$ using the measured values of $\sigma_x$, $\sigma_y$ and $\sigma_z$ to obtain $f(w_b)^{(m)}_{an}$. These are plotted as grey dotted curves for the $1/Fr > 2$ cases. The quantities $f(w_b)^{(m)}_{an}$ and $f(w_b)^{(m)}$ are almost indistinguishable for all tested cases of $1/Fr\leq4$ and for cases with $1/Fr=10$ if $St_b \leq 1$. This shows that it is appropriate to take $\sigma_x = \sigma_y = \sigma_z$ in this parameter range. Outside this range ($1/Fr < 4$), the value of the vertical component ($\sigma_z$)} becomes {noticeably} smaller than the horizontal ones  \citep{csanady_turbulent_1963,berk_dynamics_2021}. Additionally, at higher $1/Fr$, the vertical component dominates the magnitude of the bubble slip velocity, such that $\sigma_z$ contributes more to the shape of $f(w_b)$. However, {for all the cases where $1/Fr > 2$,} the anisotropy of $\sigma_i$ does not fully account for the difference from the actual $f(w_b)$. This suggests that the slip velocity components can only be considered independent when $1/Fr \leq 2$. These limitations are coincidentally compensated by the underprediction of $\sigma_i$ in our model, which reduces the width of $f(w_b)$ and leads to a satisfactory prediction of $f(w_b)$ at $1/Fr \geq 4$.

Our results hence indicate that fully capturing $f(w_b)$ is far from straightforward and would substantially add to the complexity of the model. The present approach, though relatively simple, gives a reasonable approximation of $f(w_b)$ and is therefore preferred.

\begin{figure}
  \centerline{\includegraphics{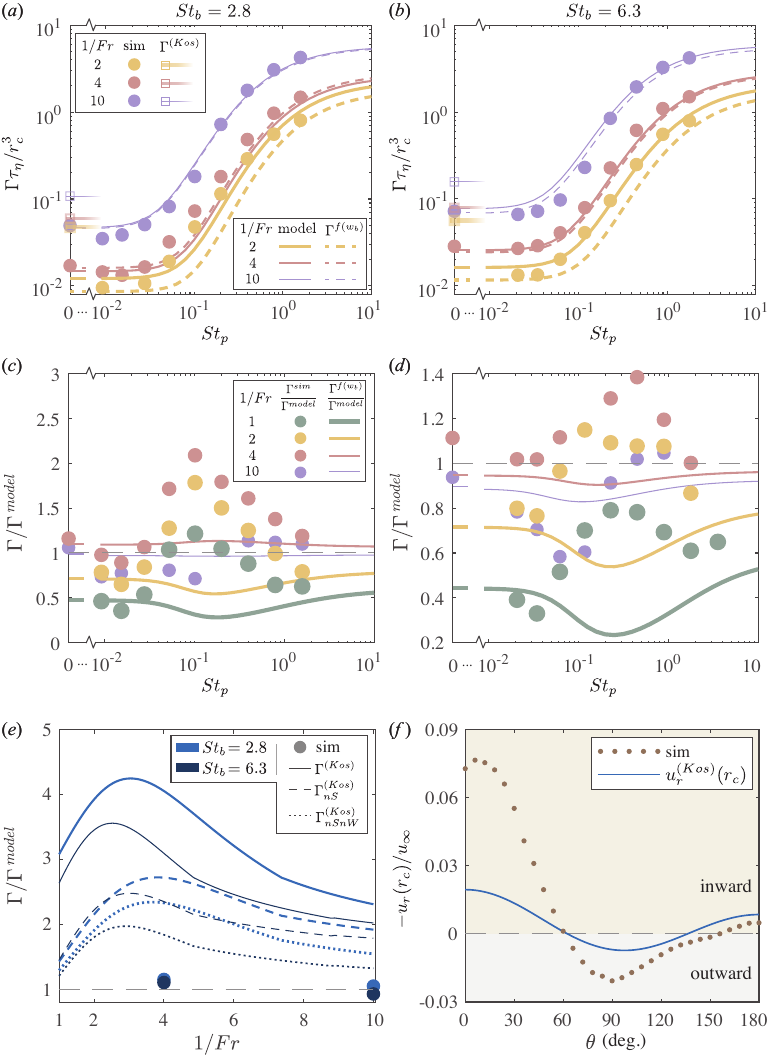}}
  \caption{(\textit{a}) The actual and predicted collision kernels at $St_b = 2.8$ and (\textit{b}) $St_b = 6.3$. Also shown are the collision kernel according to \citet{kostoglou_generalized_2020} $\Gamma^{(Kos)}$ and the collision kernel obtained from (\ref{eq:collKFromEc}) using the actual p.d.f. $\Gamma^{f(w_b)}$. The colour scheme in panel (\textit{b}) follows that of panel (\textit{a}). (\textit{c}) The ratio of the actual collision kernel $\Gamma^{sim}$ and $\Gamma^{f(w_b)}$ to the model prediction at $St_b = 2.8$ and (\textit{d}) $St_b = 6.3$. The colour scheme in panel (\textit{d}) follows that of panel (\textit{c}). (\textit{e}) Results for $\Gamma$ for $St_p=0$ and for varying $1/Fr$ relative to the model prediction. In addition to the simulation data (markers), different variants of the Kostoglou model are included as $\Gamma^{(Kos)}$ (full model), $\Gamma^{(Kos)}_{nS}$ (neglecting small-scale turbulence), and $\Gamma^{(Kos)}_{nSnW}$(neglecting small-scale turbulence and collisions in the bubble wake). (\textit{f}) The radial fluid velocity at collision distance for a stationary bubble exposed to a uniform freestream $u_{\infty}$ such that $Re_b^{\infty} = 2r_bu_{\infty}/\nu = 120$ in the simulations compared to the fit used in $\Gamma^{(Kos)}$. $-u_r(r_c)$ is plotted such that positive values imply flow towards the bubble surface.}
\label{fig:collKFS}
\end{figure}

\subsection{Collision kernel with non-settling particles} \label{sec:results_collKnonSett}
To test the effects of particle inertia and gravity independently, we initially consider  the collision kernel for the cases where the particle settling velocity is set to zero. 
We insert the model $f(w_b)$ based on (\ref{eq:meanSlip}) and $\sigma^{(BC)}$ into (\ref{eq:collKFromEc}) and compare the collision kernel predictions with the values from simulations. Figures \ref{fig:collKFS}(\textit{a} -- \textit{b}) show that the predicted and measured values of the nondimensionalised collision kernels agree reasonably well with each other for $St_b = 2.8$ and $6.3$ over the entire tested range of $St_p$ when $1/Fr \geq 2$. Consistent with the simulations, our model captures the general trend that $\Gamma\tau_\eta/r_c^3$ increases with both $St_p$ and $1/Fr$. We scrutinise this agreement by plotting the ratio of the collision kernel from simulation $\Gamma^{sim}$ to that obtained from the full model $\Gamma^{model}$ in figures \ref{fig:collKFS}(\textit{c}--\textit{d}). {The difference between $\Gamma^{sim}$ and $\Gamma^{model}$ is generally less than 30\% when $1/Fr \geq 2$ , except for $St_p \sim 0.1$ where the discrepancy becomes larger.} $St_p \sim 0.1$ corresponds to the transition regime where the interception and inertial mechanisms both play a comparable role, such that $E_c$ increases rapidly with $St_p'$ \citep{jiang_how_2024}. Since $St_p'$ is based on $w_b$, the shape of the entire distribution $f(w_b)$, not just its mean value, can influence the predicted collision kernel for intermediate values of $St_p$, unlike at extreme values of $St_p$ where the effect is less pronounced. The discrepancy between $\Gamma^{model}$ and $\Gamma^{sim}$ around $St_p = 0.1$ is hence due to the fact that our expression of $E_c$, i.e. (\ref{eq:Ec_total}), with the fitted parameters does not completely capture the interception-to-inertial transition of the collision efficiency even in a uniform background flow, as well as due to the difference between $f(w_b)$ obtained from the model and the simulations. 

We furthermore point out that $\Gamma^{sim}$ is slightly less than $\Gamma^{model}$ in most of the cases at $St_p \approx 0.01$ when $1/Fr \geq 4$, in contrast to \citet{jiang_how_2024} who found that $\Gamma^{sim}$ at $St_p = 0.01$ is larger than the predicted value obtained from (\ref{eq:collKFromEc}). Since the prediction by \citet{jiang_how_2024} is based on $f(w_b)$ measured from simulations, we plot the model prediction using the slip velocity p.d.f. from our simulations $\Gamma^{f(w_b)}$ in figures \ref{fig:collKFS}(\textit{a}--\textit{d}). The results show that the difference between $\Gamma^{model}$ and $\Gamma^{f(w_b)}$ is less than 20\% for $1/Fr \geq 4$ over all $St_p$; and $\Gamma^{f(w_b)}$ is also usually larger than $\Gamma^{sim}$ for $St_p \approx 0.01$. This rules out the modelled $f(w_b)$ as the source of the discrepancy. We also expect the frozen turbulence approximation to hold for $1/Fr \geq 4$ for these two values of $St_b$, as will be shown in \S \ref{sec:discussionValidity_FS}. We note that the simulation in \citet{jiang_how_2024} was conducted for bubbles with $(St_b,1/Fr) = (11,4.4)$ at the same $Re_\lambda$ as in this study, which translates to a larger bubble than all the cases in the present study. This may degrade the model predictions due to one or both of the following reasons. First, the measured $w_b$ may be less accurate even though the fluid velocity at the bubble position was determined with the same method (i.e. by averaging the fluid velocity within a distance of $3r_b$ around the bubble), as turbulent fluctuations over a larger area are averaged spatially. Second, the use of (\ref{eq:Ec_total}) in our model tacitly assumes that far upstream where the flow field is undistorted, the fluid velocity can be considered as uniform across the surface area bounded by the grazing trajectories (refer to figure \ref{fig:collisionsketch}). This assumption is more likely to be violated with larger bubbles in the same background turbulence. We emphasise that the above has no bearing on $\Gamma^{sim}$ because it is determined by directly counting the number of collisions. These explanations are furthermore consistent with the fact that $\Gamma^{sim}$ in the present simulations agrees very well with $\Gamma^{f(w_b)}$ when $St_p = 0$ (shown only for $1/Fr \geq 4$ in figure \ref{fig:collKFS}), where $\Gamma^{sim}$ takes a higher value than when $St_p \approx 0.01$. The slight reduction of $\Gamma^{sim}$ with increasing $St_p$ suggests that for the tested values of $Re_\lambda$, $St_b$ and $1/Fr$, turbulence activates a `negative' inertia effect at small $St_p$ because the collision angle $\theta$ defined relative to the bubble slip velocity is increased \citep{jiang_how_2024}. When the particles close to the bubble are advected by the local flow field at larger polar angles, they are slung away from the bubble by a centrifugal force. As this centrifugal force increases with particle inertia, the collision efficiency hence the collision rate in turbulence may decrease with increasing $St_p$ for particles in the very small $St_p$ regime \citep{dai_inertial_1998,huang_effect_2012}.

To put the accuracy of our model into perspective, we consider the model of \citet{kostoglou_generalized_2020}, which requires the same inputs as ours, and  include their prediction, $\Gamma^{(Kos)}$, in figures \ref{fig:collKFS}(\textit{a}--\textit{b}). Although their model was developed for particles with small $St_p$, it still overestimates the collision kernel even for tracer particles, with $\Gamma^{(Kos)}$ showing a discrepancy at least ten times larger than that of our model. In view of the excellent agreement of our model prediction with simulation data, we use our model prediction as a benchmark and plot $\Gamma^{(Kos)}/\Gamma^{model}$ for $St_p = 0$ in figure \ref{fig:collKFS}(\textit{e}). The figure shows that the overprediction persists for all $1/Fr \in [1,10]$. We note that in contrast with our model, \citet{kostoglou_generalized_2020} considers the relative motion driven by small-scale shear. This is done by adding to the (radial) collision velocity a contribution $v_{pt}$ that is uniform over all positions on the sphere with radius $r_c$ ($v_{pt} = r_p/2\cdot\langle |\partial u_i \partial x_i| \rangle = r_p/2\cdot\sqrt{2/(15\pi)}\sqrt{\varepsilon/\nu}$ in (\ref{eq:KosCollKIntegral}), where the factor 1/2 is to account only for the influx). Remarkably, despite not including this effect in our model, our predictions agree well with the data for tracer particles, suggesting that this shear mechanism likely plays a relative minor role in determining the collision kernel at the present parameters. We remove this mechanism from $\Gamma^{(Kos)}$ to obtain $\Gamma^{(Kos)}_{nS}$. As figure \ref{fig:collKFS}(\textit{e}) shows, $\Gamma^{(Kos)}_{nS} < \Gamma^{(Kos)}$ as expected, yet $\Gamma^{(Kos)}_{nS}$ still significantly and consistently overpredicts the collision rate. We therefore consider another key difference with \citet{kostoglou_generalized_2020}: the position of collisions. $\Gamma^{(Kos)}_{nS}$ includes a significant number of collisions in the bubble wake, which is attributed to the inward branches of the bubble wake vortex pair (and is enhanced by the now-neglected small-scale fluid shear effect). Such collisions are very rarely observed in the simulations in this study as well as those in \citet{jiang_how_2024} and \citet{tiedemann_direct_2025} because most of the incident particles collide with the upstream hemisphere of the bubble and are subsequently reseeded, such that the particle concentration is especially low in the bubble wake. This collision scheme corresponds to the scenario where particles attach to bubbles after collision, meaning similar behaviour might be expected in reality. Removing these collisions in the wake (i.e. setting $\theta_d = \pi$ in (\ref{eq:KosI})) gives $\Gamma^{(Kos)}_{nSnW}$ in figure \ref{fig:collKFS}(\textit{e}). Since the values of $\Gamma^{(Kos)}_{nSnW}$ are still larger than those of $\Gamma^{model}$, collisions in the bubble wake alone cannot fully account for the overprediction of the model by \citet{kostoglou_generalized_2020}. 

Taken together, this indicates that the discrepancy arises from the slip velocity magnitude and the representation of the flow distortion around the bubble in the model of \citet{kostoglou_generalized_2020}. 
Specifically, this distortion is modelled in \citet{kostoglou_generalized_2020} using a fit for the radial component of the fluid velocity adopted from \citet{nguyen_colloidal_2004} given by
\begin{equation}
    \frac{u_r^{(Kos)}(r_c)}{u_{\infty}} = -(2X\cos\theta + 3Y\cos^2\theta - Y)F,
    \label{eq:urfit}
\end{equation}
where $X$, $Y$, and $F$ are defined in (\ref{eq:KosXYZDef}), $\theta$ is the polar angle with $\theta = 0$ pointing upstream, $u_{\infty}$ is the freestream velocity, and positive values of $u_r^{(Kos)}$ correspond to flow away from the bubble. This expression is directly tested in figure \ref{fig:collKFS}(\textit{f}) for the case of a bubble in a uniform incident flow with $Re_b^{\infty} = 2r_bu_{\infty}/\nu = 120$ and $r_p/r_b = 1/30$ (see \ref{sec::schulzeFittingParameters} for simulation details). $Re_b^{\infty} = 120$ is chosen as it is the approximate value of $\langle Re_b \rangle$ for the turbulent cases with $(St_b,1/Fr) = (2.8,10)$ and $(6.3,4)$. Figure \ref{fig:collKFS}(\textit{f}) shows that $-u_r^{(Kos)}(r_c)$ slightly overpredicts the radial fluid velocity in the bubble wake, where the actual number of collisions is low due to the low particle concentration as discussed above. The radial fluid velocity on the upstream ($\theta < \ang{60}$) side of the bubble, where most of the collisions occur in the simulations, shows significant deviation between the fit and the data. 
Surprisingly, the values of $-u_r^{(Kos)}/u_{\infty}$ are much lower in this region compared to the simulation results. This implies that the overprediction of $\Gamma^{(Kos)}_{nSnW}$, at least for $(St_b,1/Fr) = (2.8,10)$ and $(6.3,4)$ at the present value of $Re_\lambda$, is driven by a significant overprediction of the freestream bubble slip velocity in turbulence, which we partly attributed further to the modelling of the vertical component in their case. There, the mean vertical slip velocity results from a simple force balance in still fluid without considering the reduction in turbulence as discussed in \S\ref{sec:model_meanVerticalSlip}. 

Two main conclusions can be drawn from the above analysis. First, $\Gamma^{(Kos)}$ consistently overestimates the collision kernel in the present parameters, partly because of collisions in the wake which become more frequent in the presence of small-scale shear. However, even when these two effects are excluded, $\Gamma^{(Kos)}_{nSnW}$ still overpredicts the collision kernel since the freestream slip velocity and the distorted flow field are not accurately modelled. For $(St_b,1/Fr) = (2.8,10)$ and $(6.3,4)$ at the present value of $Re_\lambda$, the freestream bubble slip velocity is overpredicted, though this is partially compensated by the poor agreement between the fit of the distorted flow (\ref{eq:urfit}) and the data. Therefore, caution is advised when using (\ref{eq:urfit}), as its accuracy may be limited under certain flow conditions. 
Second, our model, in spite of its simplicity, captures the essential phenomena and produces predictions that are much closer to the simulated values compared to those by \citet{kostoglou_generalized_2020}, even in the tracer particle limit.

\subsection{Collision kernel with settling particles}
\begin{figure}
  \centerline{\includegraphics{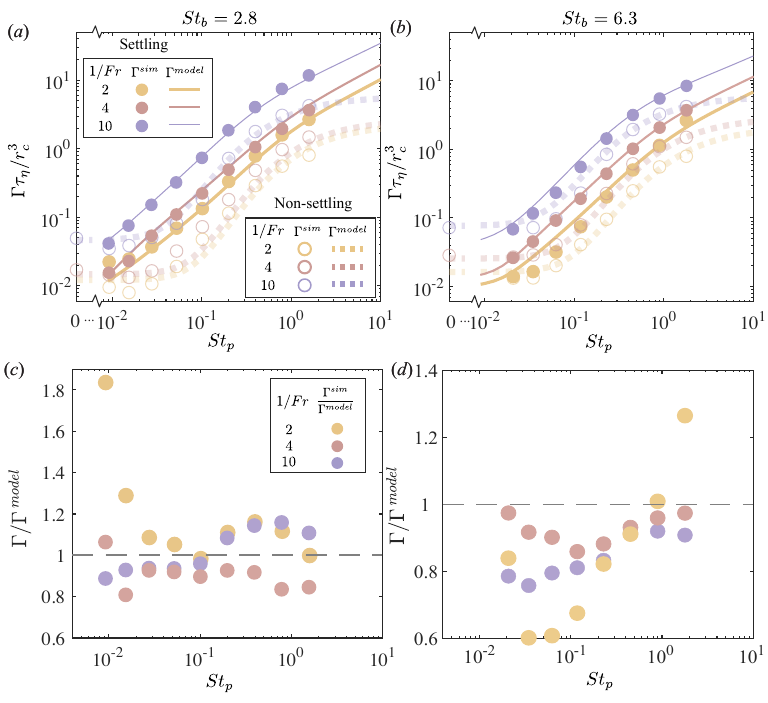}}
  \caption{(\textit{a}) The actual and predicted collision kernels at $St_b = 2.8$ and (\textit{b}) $St_b = 6.3$ with settling particles. The measured and predicted collision kernel for non-settling particles are additionally shown as open symbols and broad dotted lines, respectively. The colour scheme in panel (\textit{b}) follows that of panel (\textit{a}). (\textit{c}) The ratio of the actual collision kernel $\Gamma^{sim}$ to the model prediction at $St_b = 2.8$ and (\textit{d}) $St_b = 6.3$ for settling particles. The colour scheme in panel (\textit{d}) follows that of panel (\textit{c}).}
\label{fig:collKFS_withSettlingParticles}
\end{figure}

Now, we additionally consider the effect of particle gravity on the bubble–particle collision rate. As expected, the simulation results for the collision kernel in figure \ref{fig:collKFS_withSettlingParticles} (a–b) generally increase when particle settling is included, with the effect being most pronounced at large $St_p$. 
Generally, the simulation data is well captured by the model prediction taking gravity into account (plotted as solid lines) for both $St_b=2.8$ (in figure \ref{fig:collKFS_withSettlingParticles} (a)) and $St_b=6.3$ (figure \ref{fig:collKFS_withSettlingParticles} (b)). 
At small particle Stokes numbers, the settling is relatively weak, hence the impact of gravity-induced settling on collisions is limited in this regime. The settling velocity increases with increasing $St_p$. As a result, the collision kernel grows even more rapidly with increasing $St_p$ compared to the case without settling, implying that particle settling becomes the dominant mechanism at large $St_p$. This is particularly evident from the modelled collision kernel when nondimensionalised with $r_c^3/\tau_\eta$, which continues to increase beyond $St_p \approx 1$ when particle settling is included. 
We also point out that the difference between settling and non-settling particles is more pronounced at the lower value of $St_b$ considered here. This is because the ratio between the particle settling velocity and the bubble rising velocity becomes larger at lower $St_b$ 
(the settling velocity is solely determined by $St_p$).
Furthermore, we note that the model predictions show better agreement with the simulation data at lower $St_p$ when $1/Fr>2$, which is consistent with behaviour observed in the non-settling case.

To further assess the level of agreement, we plot in figure \ref{fig:collKFS_withSettlingParticles} (c-d) the ratio of the collision kernel obtained from simulations to that from the model. 
We note that \citet{kostoglou_generalized_2020} additionally tried to incorporate the transient misalignment between the bubble slip velocity vector and gravity through a correction coefficient $\mathfrak{f}$ (see details in \ref{sec::kostoglouExpressions}). However, adopting this coefficient did not improve the prediction and therefore this secondary effect is neglected in the present model for simplicity. 
Despite this simplification, the model prediction shows good agreement for cases of $1/Fr<2$, with deviations generally within 20\%, which verifies the correction coefficient has only a minor influence on the overall prediction.
Furthermore, the results indicate that the discrepancy between $\Gamma^{sim}$ and $\Gamma^{model}$ is reduced in the intermediate $St_p$ regime compared with the non-settling case (cf. figure \ref{fig:collKFS_withSettlingParticles}(c,d)). This is because particle settling exerts a significant influence in the intermediate $St_p$ range. 
Consequently, the deficiency associated with the modelling of $E_c^{(in)}$ in the intermediate $St_p'$ regime, as discussed in \S\ref{sec:results_collKnonSett}, becomes less significant, yielding improved agreement between the simulations and model predictions. This finding further underscores the important role of particle settling in regulating the collision process.

\section{Discussion}    \label{sec:discussion_FS}
\subsection{Model validity for the cases with lower bubble Stokes number}  \label{sec:discussionValidity_FS}
\begin{figure}
  \centerline{\includegraphics{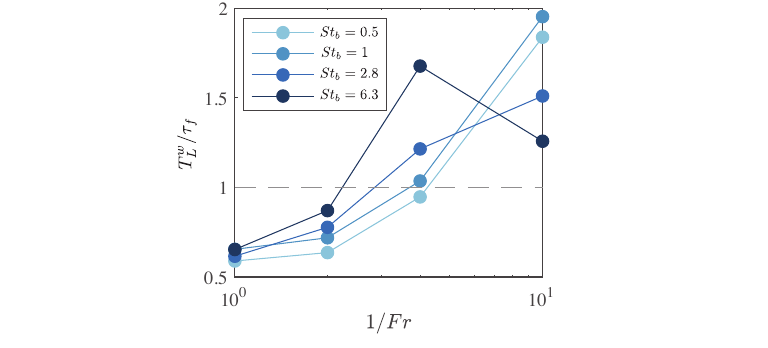}}
  \caption{The integral time scale of the bubble slip velocity.}
\label{fig:collTimeFS}
\end{figure}

In this study, we did not perform (heavy) particle-laden simulations for the cases with lower $St_b$ bubbles in view of the computation cost. Here, the validity of our model at these parameters is discussed. A key assumption in our model is the flow field around the bubble is steady during a bubble--particle collision, which can be assessed by comparing the bubble slip velocity integral time scale 
\begin{equation}
T_L^w = \int_0^{\tau_\infty} \rho_w(\tau) \mathrm{d}\tau
\end{equation}
with the fluid time scale based on the measured value of the bubble vertical velocity $\langle w_b \rangle$
\begin{equation}    \label{eq:collTimeScale}
\tau_f = \frac{2r_b}{\langle w_b \rangle},
\end{equation}
where $\rho_w(\tau) = \langle w_b(t)w_b(t+\tau) \rangle_t/\langle w_b^2(t) \rangle_t$ is the autocorrelation function of the bubble slip velocity, $\langle \cdots \rangle_t$ denotes averaging over time, and $\tau_\infty$ is defined as the value of $\tau$ where $\rho_w(\tau)$ first decays to 0.01. $T_L^w/\tau_f \gtrsim 1$ would suggest that the frozen turbulence assumption is conceptually valid and the model can be applied. 
The measured values of $T_L^w/\tau_f$ is shown in figure \ref{fig:collTimeFS}. $T_L/\tau_f$ generally increases with $1/Fr$ and is far less sensitive to $St_b$, such that for all $St_b$ considered here the frozen turbulence assumption can be applied when $1/Fr \geq 4$. For $St_b = 0.5$ and 1, we expect that modelling the slip velocity p.d.f. $f(w_b)$ by (\ref{eq:meanSlip}) and (\ref{eq:BCBubbleRMS}) might slightly underpredict the collision kernel, since the mean value of the slip velocity magnitude is underpredicted as shown in figures \ref{fig:pdfSlipVel}(\textit{a} -- \textit{b}). This discrepancy in $f(w_b)$ for $1/Fr \geq 4$ is not due to using the mean vertical rise velocity instead of the magnitude of the bubble rise velocity in the bubble drag correction factor $f_b$ in (\ref{eq:BCBubbleRMS}), because the bubble velocity at high $1/Fr$ is dominated by its vertical component. As evidence, we plot $f(w_b)$ with $\sigma_i^{(BC)}$ determined using the average $f_b$ obtained from the simulations as dashed green lines in figure \ref{fig:pdfSlipVel}. Only at small $1/Fr$ does using the actual value of $f_b$ noticeably improve the quality of the predicted slip velocity p.d.f.. If desired for these cases, a more accurate estimate of $\sigma_i^{(BC)}$ can be obtained \textit{a priori} by solving (\ref{eq:BCBubbleRMS}) iteratively using $f_{b} = 1 + 0.169(2r_b\langle w_b\rangle/\nu)^{2/3}$ as an initial guess. We additionally note that particles with $St_p \sim 1$ preferentially sample flow regions in turbulence which can reduce $\Gamma$ by up to 20\% \citep{jiang_how_2024}. Similarly preferential sampling occurs for bubbles with $St_b \sim 1$ but they cluster at different regions, which means that bubbles and particles segregate spatially when $(St_b, St_p) \sim (1,1)$ unless $1/Fr \gtrsim 10$ where segregation becomes negligible under strong gravity \citep{chan_effect_2024}. This segregation effect is not included in our model and reduces the collision kernel, hence for $(St_b, St_p) \sim (1,1)$ at intermediate values of $1/Fr$ the predicted value of the collision kernel may be closer to the actual value than anticipated.

\subsection{Implications for practice}
\subsubsection{Relevant parameter space}    \label{sec:discussion_relevantParam}
In order to evaluate the model for practically relevant conditions, we consider $r_b = 0.05\,\si{\milli\metre}$, $0.5\,\si{\milli\metre}$, and $2\,\si{\milli\metre}$ \citep{ngo-cong_isotropic_2018,ahmed_effect_1985}; $r_p \in [1, 200]\,\si{\micro\metre}$ \citep{ngo-cong_isotropic_2018}; $\varepsilon \in [0.1, 100]\,\si{W/kg}$ \citep{li_numerical_2021}; and $Re_\lambda = 100$ for sulphide minerals in water ($\rho_b = 1.2\,\si{kg/m^3}$, $\rho_p = 5000\,\si{kg/m^3}$, $\rho_f = 998\,\si{kg/m^3}$, $\nu = 1.002\times10^{-6}\,\si{m^2/s}$ and $\mathfrak{g} = 9.81\,\si{m/s^2}$). Note that here we assume $Re_\lambda = \textrm{const.}$, in contrast to \citet{ngo-cong_isotropic_2018} where $u'$ is fixed. Our choice implies that $u'$ increases with $\varepsilon$, and corresponds to the more physical scenario where a higher rotor speed drives stronger turbulence at both small and large scales. Furthermore, $Re_\lambda = 100$ was selected such that when $\varepsilon = 100\,\si{W/kg}$, then $u'^2 \approx 0.25\,\si{m^2/s^2}$, which falls within the range of values of $u'^2$ considered in \citet{kostoglou_generalized_2020}. 

\begin{figure}
  \centerline{\includegraphics{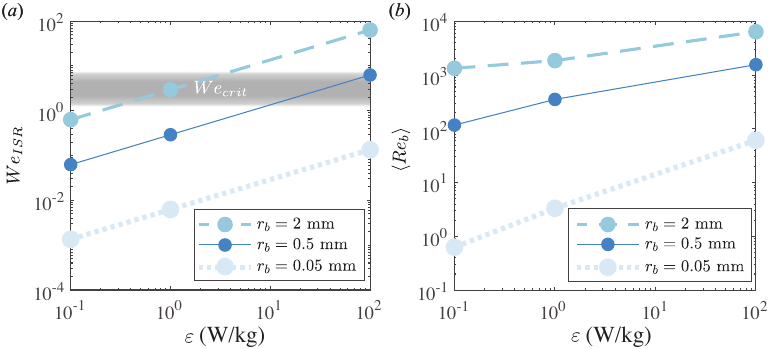}}
  \caption{(\textit{a}) The bubble Weber number based on the fluid velocity difference across the bubble. The shaded area indicates the critical Weber number ($We_{crit} \in [1.25, 7.8]$) beyond which bubbles break up. The data markers show the selected values of $\varepsilon$. Note that the lower limit of $We_{crit}$ is slightly different from the value reported by \citet{hinze_fundamentals_1955} since a coefficient of 2.13 instead of 2 is used here for the inertial subrange scaling of the second-order structure function. (\textit{b}) The predicted average bubble Reynolds number. The lines in panel (\textit{b}) are guides for the eye.}
\label{fig:WebReb}
\end{figure}

Depending on the specific value within the broad range of possible dissipation rates $\varepsilon$, the maximum stable bubble size varies because bubbles tend to break up when the local fluid shear becomes sufficiently large \citep{kolmogorov_breakage_1949,hinze_fundamentals_1955}. Quantitatively, this behaviour is captured by a critical bubble Weber number $We_{ISR} = 2.13\rho_f(2r_b\varepsilon)^{2/3}(2r_b)/\gamma$ beyond which a bubble cannot remain stable \citep{masuk_simultaneous_2021}. Here, the velocity scaling in the inertial subrange is used and $\gamma = 73\,\si{mN/m}$ is the surface tension of water. 
The nominal value for $We_{crit}$ reported in the literature vary \citep{hinze_fundamentals_1955,risso_oscillations_1998}, which is indicated in figure \ref{fig:WebReb}(\textit{a}) along with the values of $We_{ISR}$ for the combinations of $r_b$ and $\varepsilon$ investigated here. Figure \ref{fig:WebReb}(\textit{a}) shows that while most of the combinations can stably exist when accounting for the spread in $We_{crit}$, bubbles with $r_b = 2\,\si{mm}$ would break up when $\varepsilon = 100\,\si{W/kg}$. This is further consistent with the very high average bubble Reynolds number, $\langle Re_b \rangle = 2r_b\langle |\mathbf{w_b}| \rangle/\nu$, estimated by our model for this case. As shown in figure \ref{fig:WebReb}(\textit{b}), $\langle Re_b \rangle$ exceeds 6000. Based on these results, we will not evaluate our model predictions for the $(r_b,\varepsilon) = (2\,\si{mm},100\,\si{W/kg})$ case.

In addition to the physical limit imposed on $r_b$ by the background turbulence, there is an inherent maximum floatable particle size since the buoyancy of the bubble--particle aggregate must be larger than its weight for the aggregate to surface. A simple force balance gives $r_p/r_b < ((1-\rho_b/\rho_f)/(\rho_p/\rho_f - 1))^{1/3}$. This requirement is satisfied for all $r_p \leq 200\,\si{\micro\metre}$ when $r_b \geq 0.5\,\si{mm}$ and for $r_p \leq 31\,\si{\micro\metre}$ when $r_b = 0.05\,\si{mm}$ (which is indicated by open circles in the figures in \S\ref{sec:collisionRateImplications}). We acknowledge that the actual upper limit on $r_p$ is more stringent since it is also influenced by factors such as particle detachment by the background turbulence \citep{nguyen2016}. These complications are not considered here as they lie beyond the scope of this study and do not affect the discussion below.

\subsubsection{Frozen turbulence approach in practice}  \label{sec:discussion_FrozenTurbInPractice}
We now reflect on the applicability of the frozen turbulence approach at practically relevant parameters, which has a higher $Re_\lambda$ than the simulated cases. As discussed in \S\ref{sec:discussionValidity_FS}, the frozen turbulence approximation is conceptually valid when $T_L^w/\tau_f \gtrsim 1$. 
However, estimating the value of $T_L^w/\tau_f$ \textit{a priori} at a different value of $Re_\lambda$ is challenging. When $Re_\lambda$ increases, $\langle w_b \rangle$ decreases following Eq. (\ref{eq:meanSlip}) so $\tau_f$ increases according to Eq. (\ref{eq:collTimeScale}). Simultaneously, $\tau_\eta$ decreases such that $St_b$ increases, which according to our simulated cases increases $T_L^w$. The net effect of changing $Re_\lambda$ on $T_L^w/\tau_f$ is therefore uncertain.

As a crude estimate, we refer to our simulations and note that $T_L^w/\tau_f \gtrsim 1$ holds for $0.5\leq St_b \leq 6.3$ when $1/Fr \geq 4$. We furthermore note that the estimated value of $Re_\lambda\sim100$ is not orders of magnitude larger than the simulated value of $Re_\lambda = 64$. Therefore, we conjecture that for practically relevant conditions, the frozen turbulence approximation holds for $1/Fr \gtrsim O(1)$, which corresponds to $\varepsilon \gtrsim 0.2\,\si{W/kg}$. We emphasise that the above is a very rough estimate and $T_L^w/\tau_f \gtrsim 1$ is a conceptual limit. In practice, the deviation of the model prediction from reality may remain minor even when the conceptual limit is violated. This is exemplified by the $(St_b,1/Fr) = (2.8,2)$ cases in figures \ref{fig:collKFS}(\textit{c}) and \ref{fig:collKFS_withSettlingParticles}(\textit{c}), where the deviation of $\Gamma^{model}$ from $\Gamma$ is generally less than 30\% even though $T_L^w/\tau_f < 1$. In view of the uncertainty in the value $\varepsilon$ where the frozen turbulence approximation becomes invalid, we present the model predictions throughout the entire range of $\varepsilon \in [0.1,100]\,\si{W/kg}$.

\subsubsection{Collision rate predictions and implications} \label{sec:collisionRateImplications}
\begin{figure}
  \centerline{\includegraphics{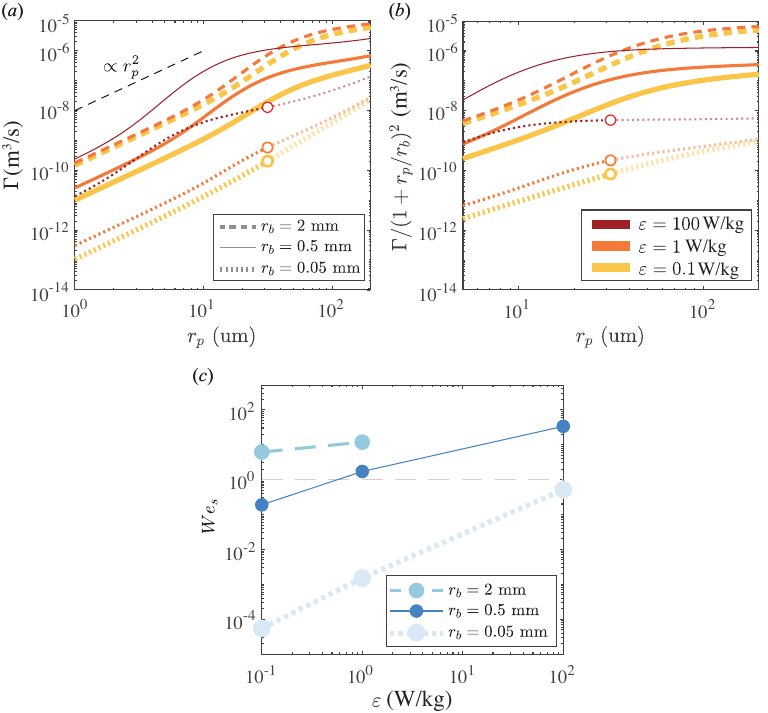}}
  \caption{(\textit{a}) The predicted collision kernel and (\textit{b}) its value when normalised by the high-inertia limit. The legends apply to both panels (\textit{a}) and (\textit{b}); the open circles indicate the maximum floatable particle size. (\textit{c}) The bubble Weber number based on the bubble slip velocity. The lines in panel (\textit{c}) are guides for the eye.}
\label{fig:predictCollK}
\end{figure}

Figure \ref{fig:predictCollK}(\textit{a}) shows that the predicted collision kernel $\Gamma$ increases monotonically with $r_p$ for the selected parameters. 
The dependence of $\Gamma$ on $r_p$ can be classified into three distinct regimes: an interception-dominated regime at small $r_p$, a transition regime at intermediate $St_p$, and a gravitational/inertia-dominated regime at large $r_p$. 
For small particles ($r_p \lesssim 7\,\si{\micro\metre}$), $\Gamma \propto r_p^2$ because the instantaneous particle Stokes number $St_p'$ is sufficiently low, such that the interception mechanism $E_c^{(i)}\propto r_p^2$ dominates (see Eq. (\ref{eq:Ec_interception})). 
In contrast, for larger particles ($r_p \gtrsim 100\,\si{\micro\metre}$), the behavior becomes more complex when examining the compensated collision kernel $\Gamma/(1 + r_p/r_b)^2$ (see figure \ref{fig:predictCollK}(\textit{b})).
In combination with larger bubbles ($r_b=0.5\,\si{mm}$ and $2\,\si{mm}$), the collision kernels scale as $(1 + r_p/r_b)^2$, corresponding to the inertial limit, which leads to the plateau observed in figure \ref{fig:predictCollK}(\textit{b}) for larger values of $r_p$.
However, for the smaller bubble ($r_b=0.5\,\si{mm}$), the collision kernel increases more rapidly than $(1 + r_p/r_b)^2$. 
This discrepancy arises from the competing contributions of particle settling and particle inertia.
For a given turbulent flow (characterized by a fixed $\varepsilon$ ), particles with the same $St_p$ exhibit identical settling velocities regardless of the bubble size. 
When the bubble size is large, the ratio between particle settling velocity and bubble rise velocity is small, such that the settling contribution to the collision kernel is weaker. Hence, the collision kernel is dominated by the particle inertia at large particle Stokes number, yielding the scaling $\Gamma \propto (1 + r_p/r_b)^2$. 
By contrast, for smaller bubble ($r_b=0.5\,\si{mm}$), the bubble rises slower and  particle settling plays a more significant role, and the compensated collision kernel continues to increase with $r_p$ in the large-$St_p$ regime, as shown in figure \ref{fig:predictCollK}(\textit{b}). 
It is also noteworthy that increasing the turbulence intensity (smaller $1/Fr$) reduces the effective particle settling velocity, thereby weakening the settling contribution. This explains the emergence of a plateau even in the case of $r_b=0.5\,\si{mm}$ and $\varepsilon=100\,\si{W/kg}$. 
Quantitatively, increasing $r_b$ leads to larger values of $\Gamma$, partly because $\Gamma \propto r_b^2$ by geometry.

Higher values of $\Gamma$ are also predicted by our model for larger values of $\varepsilon$. The increase in $\Gamma$ is driven by multiple factors since $f(w_b)$ as well as $Re_b$, and hence also $St_p'$ all vary. Furthermore, the boundaries of the transition regime clearly shifts towards smaller values of $r_p$ at larger values of $\varepsilon$. This is because $\varepsilon$ and $\langle|\mathbf{w_b}|\rangle$ are positively correlated for the parameters considered, which can be inferred from the increase of $\langle Re_b \rangle = 2r_b\langle |\mathbf{w_b}| \rangle/\nu$ with $\varepsilon$ in figure \ref{fig:WebReb}(\textit{b}) (note that the increase is driven by enhanced turbulent fluctuations of the bubble slip velocity, while the mean vertical slip velocity $\langle w_b\rangle$ is reduced in accordance with (\ref{eq:meanSlip})). Therefore, particles of a particular size would have a larger $St_p'$ on average for larger values of $\varepsilon$, such that the inertial mechanism is activated and the fully inertial limit is reached at lower values of $r_p$. Additionally, this increase in $\langle |\mathbf{w_b}| \rangle$ raises the associated bubble Weber number defined with the bubble slip velocity $We_s = 2\rho_f r_b \langle |\mathbf{w_b}| \rangle^2/\gamma$. As figure \ref{fig:predictCollK}(\textit{c}) shows, $We_s \gtrsim 1$ for $\varepsilon \gtrsim 1\,\si{W/kg}$ when $r_b \gtrsim 0.5\,\si{mm}$, which suggests that the spherical bubble assumption may no longer hold, hence the quantitative predictions in this parameter range should be used with caution. Nonetheless, deviations from the spherical bubble assumption are likely higher-order effects. In general, our model prediction at typical flotation parameters indicate that inertial effects are not negligible and need to be taken into account. This is true even for $r_p = 7\,\si{\micro\metre}$ which is encountered in fine particle flotation \citep{lynch_mineral_1981} as long as $\varepsilon$ is sufficiently large ($\varepsilon \gtrsim 5\,\si{W/kg}$ for $r_b = 0.05\,\si{mm}$).

\begin{figure}
  \centerline{\includegraphics{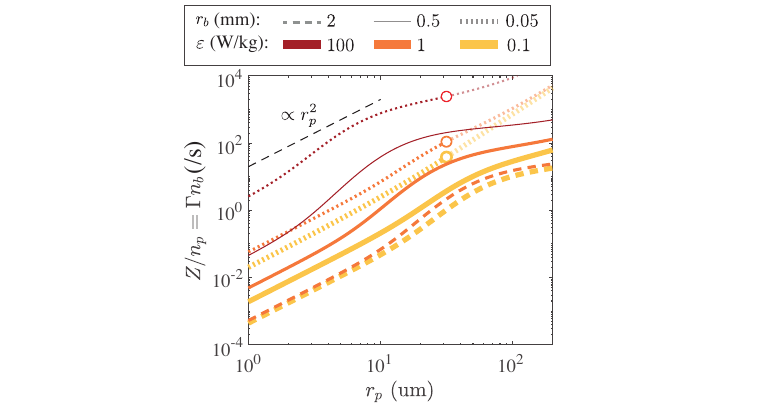}}
  \caption{The predicted collision rate per particle when the gas holdup is fixed at 10\%. The open circles indicate the maximum floatable particle size.}
\label{fig:predictCollK_constAlpha}
\end{figure}

While the collision kernel facilitates comparison between scenarios with different bubble and particle number densities, it is not the most direct comparison in practice when $r_b$ varies since it implies that the gas holdup $\phi \propto r_b^3$, which may not be the case \citep{yianatos_gas_2010}. This is an important detail, especially for the flotation of fine particles. In this context, the mineral recovery rate is low due to reduced collision efficiency at small $r_p$, so smaller bubbles are employed to mitigate this problem \citep{ahmed_effect_1985,miettinen_limits_2010}. The collision kernel then takes a lower value as has already been observed in figure \ref{fig:predictCollK}(\textit{a}). However, considering only $\Gamma$ in this scenario ignores the positive effect on the overall collision rate due to the increase in the bubble number density assuming a constant gas volume flow rate. Therefore, in figure \ref{fig:predictCollK_constAlpha} we plot $\Gamma n_b$ (i.e. the collision rate per particle) for constant $\varepsilon_g$ at a typical value of 10\% \citep{yianatos_gas_2010}. As expected, the shape of the curves are identical to those in figure \ref{fig:predictCollK}(\textit{a}); though crucially, $\Gamma n_b$ increases when $r_b$ decreases. This shows that using smaller bubbles has a net positive effect on the collision rate, and our model prediction supports the usage of smaller bubbles for flotation, which is consistent with results in the literature \citep{ahmed_effect_1985,miettinen_limits_2010}.

\section{Conclusion}\label{sec:conclusion_FS}
\begin{figure}
  \centerline{\includegraphics[width=\linewidth]{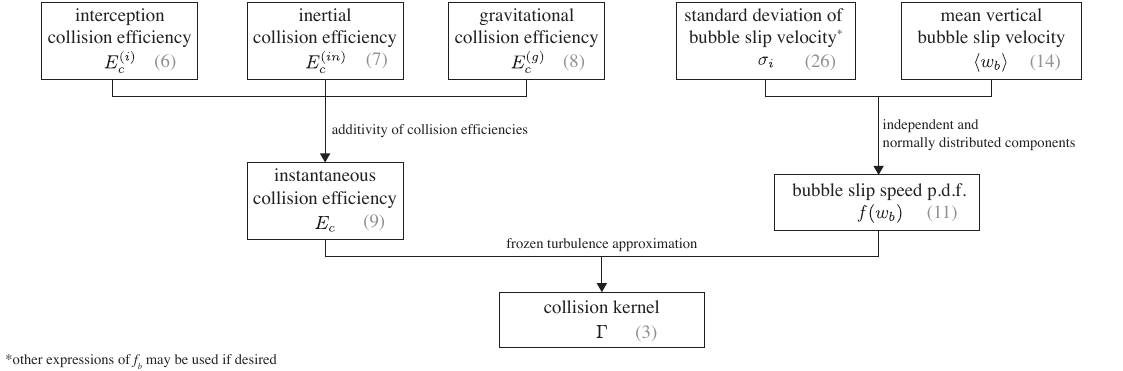}}
  \caption{{Summary of the proposed model. The numbers in brackets refer to the expression of the corresponding quantities.}}
\label{fig:modelOverview}
\end{figure}

In this study, we develop a model that predicts \textit{a priori} the bubble--particle collision rate of inertial particles in homogeneous isotropic turbulence given the bubble, particle, and turbulence properties: namely, the bubble radius, the bubble density, the particle radius, the particle response time, the liquid kinematic viscosity, the liquid density, the Taylor Reynolds number, the dissipation rate, and gravity. Our model accounts for bubble behaviour in turbulence as well as the distorted flow field around finite-size bubbles. A summary of our model is displayed in figure \ref{fig:modelOverview}. As figure \ref{fig:modelOverview} shows, our model adopts the frozen turbulence approximation proposed by \citet{jiang_how_2024}, such that the existing parametrisations of the bubble--particle collision efficiency in still fluid can be applied.
The main input required for this approach is the bubble slip speed p.d.f.. We assume independent bubble slip velocity components and model their probability density functions using normal distributions with the mean of the vertical component given by \citet{ruth_effect_2021} and \citet{liu_direct_2024} (i.e. (\ref{eq:meanSlip})), and the standard deviation given by \citet{berk_analytical_2024} (i.e. (\ref{eq:BCBubbleRMS})). The resulting bubble slip speed p.d.f.s are compared with finite-size bubble simulations over $1\leq 1/Fr\leq 10$ with $0.5 \leq St_b \leq 6.3$ and $Re_\lambda = 64$. Good agreement is observed for $1/Fr \geq 4$. For $St_b = 2.8$ and $6.3$, we additionally calculate the bubble--particle collision rates from the simulations and find satisfactory agreement with the model predictions. This result is consistent with our proposed requirement that $T_L^w/\tau_f \gtrsim 1$ for the frozen turbulence approximation, which suggests that the model should be valid for $1/Fr \geq 4$ when $0.5 \leq St_b \leq 6.3$ and $Re_\lambda = 64$. Applying our model to typical flotation parameters confirms that the collision rate can be raised by using larger particles, smaller bubbles, or stronger turbulence. Importantly, we find that even for fine particles, particle inertia can play a significant role in determining the overall collision rate.

In summary, our model includes the particle inertia and particle settling effects on the collision rate and provides significantly better predictions of the bubble--particle collision rate in homogeneous isotropic turbulence than existing theories, even at the $St_p \to 0$ limit. Considering the high inhomogeneity of the turbulence in actual flotation cells, our model can be used on a subgrid-scale level in industrial computer fluid dynamics simulations to predict the local collision rate, at least for smaller bubble and particle Stokes numbers where the behaviour of the dispersed phases is determined by the local flow field. 
Presently, the expressions for the collision efficiency and the mean vertical slip velocity presented in (\ref{eq:Ec_total}) and (\ref{eq:meanSlip}) are for fully contaminated bubbles. In principle though, our general modelling approach (i.e. the frozen turbulence approach and the modelling of the bubble slip speed p.d.f. with independent normally distributed components) is not restricted to such bubbles, as long as the corresponding expressions of the collision efficiency and the mean vertical slip velocity are used. 
Potential future improvements of the model include a more accurate representation of $E_c^{(in)}(St_p)$, to more faithfully capture the dependence of the collision rate on $St_p$; 
as well as incorporating the turbulence-induced spatial segregation of bubbles and particles \citep{calzavarini_quantifying_2008-1,chan_bubbleparticle_2023,chan_effect_2024}, which can reduce the collision rate especially when $St_p \approx 1$ \citep{jiang_how_2024}. 
In addition, interactions with neighbouring bubbles may influence the collision kernel in practice, but these effects are not currently represented in the model and should be addressed in future research.

\setlength{\nomitemsep}{-4pt}
\nomenclature{$\Gamma$}{collision kernel}
\nomenclature{$Z$}{collision rate per unit volume}
\nomenclature{$n_p$}{particle number density}
\nomenclature{$n_b$}{bubble number density}
\nomenclature{$r_g$}{grazing radius}
\nomenclature{$r_p$}{particle radius}
\nomenclature{$r_b$}{bubble radius}
\nomenclature{$E_c$}{collision efficiency in still fluid}
\nomenclature{$n_b$}{bubble number density}
\nomenclature{$w_b$}{bubble slip velocity magnitude}
\nomenclature{$Re_b$}{bubble Reynolds number based on $w_b$}
\nomenclature{$St_p'$}{particle Stokes number based on time scale $2r_b/w_b$}
\nomenclature{$v_s$}{particle settling velocity}
\nomenclature{$\theta_{cap}$}{contamination angle}
\nomenclature{$\tau_p$}{particle response time scale}
\nomenclature{$\eta$}{Kolmogorov length scale}
\nomenclature{$\rho_p$}{particle density}
\nomenclature{$\rho_f$}{fluid density}
\nomenclature{$E_c^{(i)}$}{interceptional collision efficiency}
\nomenclature{$E_c^{(in)}$}{inertial collision efficiency}
\nomenclature{$Re_{\lambda}$}{Taylor Reynolds number}
\nomenclature{$\nu$}{fluid kinematic viscosity}
\nomenclature{$\varepsilon$}{turbulent energy dissipation rate}
\nomenclature{$Fr$}{Froude number}
\nomenclature{$\tau_\eta$}{Kolmogorov time scale}
\nomenclature{$St_p$}{particle Stokes number based on $\tau_\eta$}
\nomenclature{$\tau_b$}{bubble response time scale}
\nomenclature{$u_\eta$}{Kolmogorov velocity scale}
\nomenclature{$w_{b,i}$}{bubble slip velocity component in $i$-th direction}
\nomenclature{$\boldsymbol{v}_b$}{bubble rise velocity}
\nomenclature{$\boldsymbol{\Tilde{v}_b}$}{bubble velocity fluctuation}
\nomenclature{$\boldsymbol{u}$}{fluid velocity}
\nomenclature{$\Tilde{\boldsymbol{u}}$}{fluid velocity fluctuation}
\nomenclature{$v_q$}{bubble rise velocity in still fluid}
\nomenclature{$\sigma$}{standard deviation of the bubble slip velocity}
\nomenclature{$E(\omega)$}{Energy spectrum}
\nomenclature{$We$}{Weber number}
\nomenclature{$T^w_L$}{integral time scale of bubble slip velocity}
\nomenclature{$\rho_w$}{autocorrelation function of the bubble slip velocity}

\begin{tcolorbox}[colframe=black, colback=white, sharp corners, boxrule=0.5pt]
\begin{multicols}{2}
\printnomenclature
\end{multicols}
\end{tcolorbox}

\section*{Funding}
This project has received funding from the European Research Council (ERC) under the European Union's Horizon 2020 research and innovation programme (grant agreement No. 950111, BU-PACT). We acknowledge the EuroHPC Joint Undertaking for awarding the project EHPC-REG-2023R03-178 access to the EuroHPC supercomputer Discoverer, hosted by Sofia Tech Park (Bulgaria).

\section*{CRediT authorship contribution statement}
Timothy T.K. Chan: methodology, formal analysis, software, data curation, writing – original draft, writing – review \& editing, visualization. 
Linfeng Jiang: methodology, software, validation, formal analysis, investigation, data curation, writing – original draft, writing - review \& editing.
Dominik Krug: conceptualization, methodology, resources, writing - review \& editing, supervision, project administration, funding acquisition.

\section*{Declaration of competing interest}
The authors declare that they have no known competing financial interests or personal relationships that could have appeared to influence the work reported in this paper.

\section*{Data availability statement}
All data supporting this study are available from the authors upon request.

\appendix
\section{Fitting parameters for the inertial collision efficiency}    \label{sec::schulzeFittingParameters}
\begin{figure}
  \centerline{\includegraphics{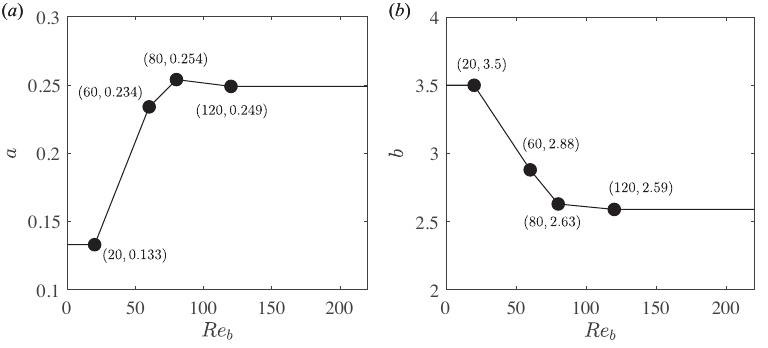}}
  \caption{The value of (\textit{a}) $a$ and (\textit{b}) $b$ in the expression of $E_c^{(in)}$, i.e. (\ref{eq:Ec_total}), as a function of $Re_b$.}
  \label{fig:Ec_fitParam}
\end{figure}

We determine the fitting parameters $a$ and $b$, which are used in the expression of $E_c^{(in)}$ (\ref{eq:Ec_inertial}) from supplementary simulations of a fixed bubble and freely-moving point-particles in a fluid without external forcing. We use the same fluid and point-particle solvers as described in \S\ref{sec:methods_eq_FS} with $\mathbf{f} = \mathbf{0}$ in (\ref{eq:N-S}). In contrast to the simulations discussed in the main text, the bubble is always fixed in the centre of the computation domain, whose size is $512\times512\times768$ and is periodic along the two horizontal directions. On the top boundary, a uniform downward fluid velocity is imposed to give the desired $Re_b$ and $\sim 10000$ particles with $r_p = r_b/30$ are injected initially above the bubble. Depending on the location of the particle in relation to the grazing trajectory, it either collides with the bubble and is removed immediately, or is advected by the flow and eventually exits the bottom of the domain, where the longitudinal fluid velocity gradient is set to be uniformly zero. $r_b$ is resolved by 40 grids, which satisfies both the boundary layer and particle resolution requirements discussed in \S\ref{sec:methods_sim_FS} for $Re_b\in[20,120]$. The values of the fitting parameters $a$ and $b$ at $Re_b = 20,60,80$, and 120 are shown in figure \ref{fig:Ec_fitParam}. Between these values of $Re_b$, the fitting parameters are linearly interpolated. Beyond $Re_b \in [20,120]$, we take $(a,b) = (0.133,3.5)$ for $Re_b < 20$, and $(a,b) = (0.249,2.59)$ for $Re_b > 120$ as the curves flatten at high $Re_b$.

\section{Expressions of \citet{kostoglou_generalized_2020}} \label{sec::kostoglouExpressions}
In this appendix, we provide the expressions used to evaluate $\Gamma^{(Kos)}$ for completeness. 
This model is based on the combined surface and velocity integral 
\begin{eqnarray}    \label{eq:KosCollKIntegral}
    \Gamma^{(Kos)} &=& \oiint_{S_c} \iiint\limits_{-\infty}^\infty \int_{0}^{\infty} 
    \Lambda(-\Delta v_r\cdot f(w_{bo,x})f(w_{bo,y})f(w_{bo,z})f(v_{pt})\\ \nonumber
    &&\mathrm{d}v_{pt}\mathrm{d}w_{bo,x}\mathrm{d}w_{bo,y}\mathrm{d}w_{bo,z})\mathrm{d}S,
\end{eqnarray}
where $S_c$ is a sphere with a radius of $r_b + r_p$, $\Delta v_r$ is the radial component of the bubble--particle relative velocity, which is positive when the pair is separating,
\begin{equation}    \label{eq:KostoglouBestFit}
\Lambda(x) = 
\left\{
\begin{aligned}
x     \quad&\text{for $x > 0$,}\\
0 \quad&\text{otherwise,}\\
\end{aligned}
\right.
\end{equation}
$f(\cdot)$ is the probability density function (p.d.f.), $w_{bo,x,y,z}$ are components of the bubble slip velocity with respect to the fluid velocity in the absence of the bubble, and $v_{pt}$ is the relative bubble--particle velocity due to local velocity gradients.  
Rewriting (\ref{eq:KosCollKIntegral}),
\begin{eqnarray}
    \Gamma^{(Kos)} =2\pi(r_b + r_p)^2 I,
\end{eqnarray}
where
\begin{equation}    \label{eq:KosI}
    I = N_1 \frac{2 - \cos^3\theta_c + \cos^3\theta_d}{3} + N_2 \frac{-\cos^2\theta_c + \cos^2\theta_d}{2} + N_3(2 - \cos\theta_c + \cos\theta_d).
\end{equation}
$N_{1,2,3}$ and $\theta_{c,d}$ will be defined in \S\ref{subsec::kostoglouDef_N} and \S\ref{subsec::kostoglouDef_theta}, respectively.

\subsection{Definition of $N_{1,2,3}$}  \label{subsec::kostoglouDef_N}
$N_{1,2,3}$ characterise the magnitude of the radial component of the bubble--particle relative velocity in the bubble reference frame and are defined as
\begin{equation}
    N_1 = 3YFU_T,\quad N_2 = 2XFU_T - v_s\mathfrak{f},\quad N_3 = v_{pt} - YFU_T.
\end{equation}
Here, $v_s = 2r_p^2(1 - \rho_p/\rho_f)\mathfrak{g}/(9\nu (1+0.169Re_p^{2/3}))$ is the particle terminal velocity with non-Stokesian drag correction, $v_{pt} = (r_p/\tau_\eta)/(30\pi)^{1/2}$ is the collision velocity attributed to local fluid shear,
\begin{equation}
\frac{U_T}{\sigma_i} = 
\left\{
\begin{aligned}
1.6,     \quad&\text{for $\alpha < 0.1$}\\
-0.0188\alpha^3+0.2174\alpha^2+0.1073\alpha+1.5552,\quad&\text{for $0.1\leq \alpha\leq 5$}\\
\alpha + (1/\alpha),\quad&\text{for $\alpha>5$}
\end{aligned}
\right.
\end{equation}
from \citet{chan_effect_2024}, and $\alpha = v_q/\sigma_i$ where $v_q$ is the bubble rise velocity in still fluid. Taking $f_{b} = 0.426Re_b^{1/2}$,
\begin{equation}
    v_q = \Bigg[0.261 \sqrt{\frac{2r_b^3}{\nu}} \bigg(1 - \frac{\rho_b}{\rho_f}\bigg)\mathfrak{g} \Bigg]^{2/3}
\end{equation}
and
\begin{equation}    \label{eq:KosCorrectedRMS_fbsqrtRe}
    \sigma_i = 2u'\bigg(1 + \frac{3.79\nu^{1/2}u'^2\sigma_i^{1/2}}{\varepsilon r_b^{3/2}}\bigg)^{-1/2}.
\end{equation}
Note that the coefficient in the numerator of (\ref{eq:KosCorrectedRMS_fbsqrtRe}) is $3.79$ instead of $2.625$ as presented in \citet{kostoglou_generalized_2020} due to a typographical error. $X$, $Y$, $F$ are related to the bubble flow field in still fluid and are given by
\begin{equation}    \label{eq:KosXYZDef}
    X = \frac{3}{2} + \frac{9}{32}\frac{Re_b}{(1 + 0.31Re_b^{0.7})},\quad Y = \frac{3}{8}\frac{Re_b}{(1 + 0.217Re_b^{0.518})},\quad F = \frac{1}{2}\bigg(\frac{r_p}{r_b}\bigg)^2
\end{equation}
where $Re_b = 2r_b U_T/\nu$ in (\ref{eq:KosXYZDef}) only.
Finally, $\mathfrak{f}$ describes the angle between the bubble slip velocity with gravity and
\begin{align}
    \mathfrak{f} &=
		\begin{cases}
			\, \alpha/2 &\text{for $\alpha < 1$} \\
			\, 0.5 + 0.5[1 - \exp(-0.85(\alpha - 1))] &\text{for $\alpha\geq1$.}
		\end{cases}
\end{align}

\subsection{Definition of $\theta_{c,d}$}     \label{subsec::kostoglouDef_theta}
The radial component of the bubble--particle relative velocity $\Delta v_r$ has an angular dependence on the polar angle $\theta$, which is defined such that $\theta = 0$ points towards the incident flow. Depending on the bubble properties, $\Delta v_r$ may change signs. The corresponding critical angles are given by $\theta_c$ and $\theta_d$, such that $\Delta v_r$ is positive (velocity pointing radially outwards) when the polar angle $\theta\in(\theta_c, \theta_d)$. Generally,
\begin{equation}
    \cos\theta_c = \frac{-N_2 + \sqrt{N_2^2 - 4N_1N_3}}{2N_1},\quad \cos\theta_d = \frac{-N_2 - \sqrt{N_2^2 - 4N_1N_3}}{2N_1}.
\end{equation}
If an imaginary value or $|\cos\theta_{c,d}| > 1$ is obtained, the corresponding solution is rejected using the following procedure. In the case where both $\theta_c$ and $\theta_d$ are rejected, $\theta_c = \theta_d = \pi$. Otherwise, if the solution for $\theta_c$ is rejected, $\theta_c = 0$; while if the solution for $\theta_d$ is rejected, $\theta_d = \pi$.

\section{Comparison between the full and approximate expressions of $f(w_b)$}    \label{sec::slipVelNormDist}
\begin{figure}
  \centerline{\includegraphics{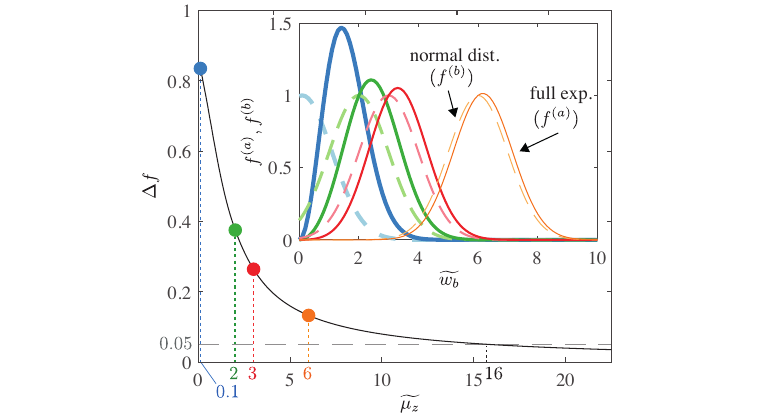}}
  \caption{The discrepancy between $f(w_b)$ modelled by integrating the joint p.d.f. of the individual bubble slip velocity components and by a normal distribution. The inset shows the distributions resulting from these two models at various $\widetilde{\mu_z}$, the values of which are given by the data points with the same colour in the main panel.}
  \label{fig:normDistvsFull}
\end{figure}

At large $\langle w_b \rangle$, the complete expression of $f(w_b)$ reported in (\ref{eq:pdfSlipVel}) converges to a normal distribution as given in (\ref{eq:pdfSlipVelComplete}). To compare  the two expressions in (\ref{eq:pdfSlipVelComplete}), we rewrite them in terms of the dimensionless variables $\widetilde{w_b} = w_b/\sigma$ and $\widetilde{\mu_z} = \langle w_b \rangle/\sigma$. This yields
\begin{equation}
	f(w_b) = \frac{\widetilde{w_b}^2}{\sqrt{2\pi} \sigma}\exp{\bigg(-\frac{\widetilde{w_b}^2 + \widetilde{\mu_z}^2}{2} \bigg)}\bigg(\frac{2}{\widetilde{w_b}\widetilde{\mu_z}}\sinh{(\widetilde{w_b} \widetilde{\mu_z})}\bigg) = \frac{1}{\sqrt{2\pi} \sigma}f^{(a)}
\end{equation}
and
\begin{equation}
	f(w_b) = \frac{1}{\sqrt{2\pi}\sigma}\exp\bigg(-\frac{(\widetilde{w_b} - \widetilde{\mu_z})^2}{2}\bigg)  = \frac{1}{\sqrt{2\pi} \sigma}f^{(b)}.
\end{equation}
We therefore plot $f^{(a)}$ and $f^{(b)}$ in the inset of figure \ref{fig:normDistvsFull}, which shows that the two curves converge at large $\widetilde{\mu_z}$. To quantify this, the main panel of figure \ref{fig:normDistvsFull} displays
\begin{equation}
    \Delta f = \frac{\int_0^\infty |f^{(a)} - f^{(b)}|\mathrm{d}\widetilde{w_z}}{\int_0^\infty f^{(a)}\mathrm{d}\widetilde{w_z}} = \frac{\int_0^\infty |f^{(a)} - f^{(b)}|\mathrm{d}\widetilde{w_z}}{\sqrt{2\pi}}
\end{equation}
over a range of $\widetilde{\mu_z}$. 
$\Delta f$ decreases with increasing $\widetilde{\mu_z}$ consistent with our previous observation, and drops below $0.05$ when $\widetilde{\mu_z} > 16$. We hence choose $\widetilde{\mu_z} = 16$ as the transition value between the two expressions shown in (\ref{eq:pdfSlipVelComplete}).

\section{Grid convergence test} \label{sec::GridConvergence}
\begin{figure}      
  \centerline{\includegraphics{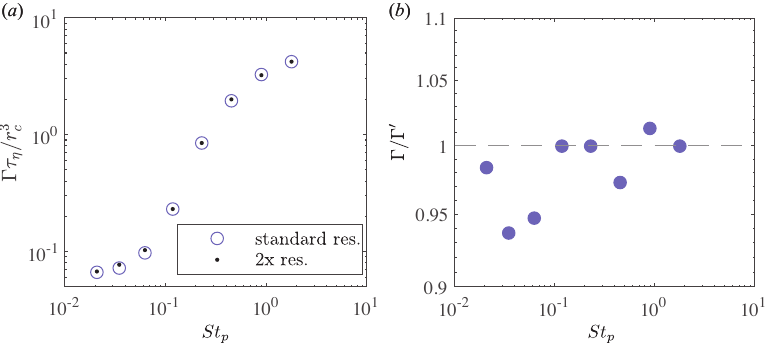}}
  \caption{(\textit{a}) The collision kernels at the standard grid resolution $\Gamma$ and that at the doubled grid resolution $\Gamma'$, and (\textit{b}) the ratio of $\Gamma/\Gamma'$ as a function of the particle Stokes number for $(St_b, 1/Fr) = (6.3,10)$.}
  \label{fig:collKResTest}
\end{figure}

To ensure that the collision kernel is not sensitive to the grid spacing, we perform an additional simulation for the highest $Re_b$ case ($(St_b, 1/Fr) = (6.3,10)$) at twice the grid resolution for non-settling particles. Figure \ref{fig:collKResTest} shows that the ratio of the collision kernels obtained is almost unity, which suggests that the grid resolution is sufficient.

\bibliographystyle{elsarticle-harv}

\end{document}